\documentclass[pre,amsmath,onecolumn]{revtex4-2}
\usepackage[normalem]{ulem}
\usepackage{amsfonts} 
\usepackage{amsmath} 
\usepackage{color} 
\usepackage{hyperref}
\setcitestyle{square}
\usepackage{subfigure} 
\usepackage{natbib} 
\usepackage{graphicx,amssymb}

\usepackage{gensymb} 
\usepackage{mathtools} 

\setcitestyle{square}

\usepackage{enumitem}

\begin{document} 

\title{Anomalous diffusion in the citation time series of scientific publications}
\author{Maryam Zamani\textsuperscript{a},  Erez Aghion\textsuperscript{a}, Peter Pollner\textsuperscript{b}, Tamas Vicsek\textsuperscript{b}, Holger Kantz\textsuperscript{a}}
 \affiliation{\makebox[\textwidth][c]{a) Max-Planck Institute for the Physics of Complex Systems, Dresden D-01187, Germany}  \\{b) Department of Biological Physics, E\"otv\"os University, Budapest, 1117, Hungary} }
%%%%%%%%%%%%%%%%%%%%%%%%%%%%%%%%%%%%%%%%%%%%%%%%%%%%%%%%%%%%%%%%%%%%%%%%%%%%%%%
%
% A B S T R A C T 
%
%%%%%%%%%%%%%%%%%%%%%%%%%%%%%%%%%%%%%%%%%%%%%%%%%%%%%%%%%%%%%%%%%%%%%%%%%%%%%%%
\begin{abstract}
We analyze the citation time-series of manuscripts in three different fields of science;  physics, social science and technology. The evolution of the time-series  of the yearly number of citations, namely the citation trajectories, diffuse anomalously, their variance scales with time $\propto t^{2H}$, where $H\neq 1/2$. We provide detailed analysis of the various factors that lead to the anomalous behavior: non-stationarity, long-ranged correlations and a fat-tailed increment distribution. The papers exhibit high degree of heterogeneity, across the various fields, as the statistics of the highest cited papers is fundamentally different from that of the lower ones. The citation data is shown to be highly correlated and non-stationary; as all the papers except the small percentage of them with high number of citations, die out in time. 
\end{abstract}

\maketitle{}

\section{Introduction} 
\label{Introduction}

Studying the structure of underlying patterns in different fields of science gives insight into the evolutionary process of the system over time, and also provides the ability to make general assumptions about its future \cite{kuhn2012structure,evans2013future,king2009automation}. 
Citation networks can be used for not only revealing the hidden patterns and structure in different fields of science, but they also disclose the substantial human behavior which forms them  \cite{wu2019large,deville2014career,sinatra2016quantifying,sekara2018chaperone}.   
Following a growing interest, the number of publications on citation analysis has been significantly increasing in the past few decades \cite{fortunato2018science,zamani2020evolution,garfield1955citation}. 
The analysis so far suggests a nearly universal behavior in science across very different disciplines \cite{radicchi2008universality,wang2013quantifying,redner2005citation,redner1998popular,uzzi2013atypical}.
One prominent observation reveals the presence of a small percentage of papers in each discipline that dominate the field, with the number of their citations growing increasingly faster as they gain more attention in time, whereas on the other hand there is a large number of papers whose influences are quickly dying out \cite{yin2017time,golosovsky2017power,stringer2010statistical,barabasi2012handful}. This heterogeneity, as we will also show, is related to the anomalous growth of the variance of the process. \\
In this paper, we study the yearly citation time series in three different fields: physics, social science and technology. The records have been collected from the WoS database \cite{ISI}, and include publications between the years 1974 and 2012. Within this dataset, we chose papers which were published during five years, 1974 to 1978, and track their citation time series up until 34 years after publication.  We use two different methods to calculate the temporal behavior of the average yearly number of citations, as well as their variance, and we study the non-stationarity of the time series, its correlations and the effect of large fluctuations. 

In the Time-Average (TA) approach, we first consider the quantity $C_i(t')$, defined as the number of citations that an individual paper $i$ gains in a single year, $t'$ years after its original publication. Then, the average citations number for this paper after $t$ years, is $\overline{C_i(t)}=\frac{1}{t}\sum_{t'=1}^t C_i(t')$. In the Ensemble-Average (EA) method, we look at the number of citations of each of the manuscripts per year, without considering their history. Thus, the ensemble mean at year $t$ is $\langle C(t)\rangle=\frac{1}{N}\sum_{i=0}^N C_i(t)$. Here, $N$ is the total number of analysed papers. It is not uncommon practice to combine the two former methods, and obtain the Ensemble-Average Time Average (EATA); $\langle \overline{C_{i}(t)}\rangle=\frac{1}{N}\sum_{i=0}^{N}\frac{1}{t}\sum_{t'=1}^{t}C_{i}(t')$. According to the law of large numbers, if $N$ is large and the system is in statistical equilibrium, the two approaches yield completely similar results: the probability distribution of the TAs becomes a $\delta$-function around its mean, and $\langle\overline{C_{i}(t)}\rangle=\langle{C(t)}\rangle$. When this result is violated, however, the system is said to exhibit weak ergodicity breaking \cite{bouchaud1992weak,bel2005weak,burov2011single}. In such a case it is important to discuss the difference between the two approaches, as mean values in the system depend on the method of measurement. Note that weak ergodicity breaking is more commonly discussed in the context of the mean squared displacement versus time-averaged mean squared displacement (see definitions, below), which describes the time evolution of the increment fluctuations in the data, see e.g., \cite{cherstvy2013anomalous,thiel2014weak,metzler2015weak,manzo2015weak}. 
Using both approaches for calculating the mean statistical properties of the system, and using data from three different disciplines, we found that the citation trajectories diffuse anomaly and are non-ergodic. Therefore, we investigate the  anomalous diffusion of citations by a detailed analysis to find the basic factors which lead to such anomalous behavior. First we study the correlation of citations and quantify it by the so-called Joseph exponent \cite{mandelbrot1968noah,chen2017anomalous}, which has been derived from time-average of the Mean Squared Displacement (MSD). The second and third factors which lead to anomalous growth of citation time series are non-stationarity, quantified by the Moses effect, \cite{chen2017anomalous,meyer2018anomalous,aghion2020moses} and the Noah effect \cite{mandelbrot1968noah,chen2017anomalous,meyer2018anomalous,aghion2020moses} (stemming from fat-tailed distribution) which is quantified using Latent exponent.

The structure of the paper is as follows: In sub-Sec. \ref{Citationdistribution}, we study the yearly citation distributions and compare them for three different disciplines. In sub-Sec.  \ref{MeanVariance}, we investigate the mean and variance of the yearly citation number, comparing time- and ensemble-averages, for all the papers in the field, as well as when we separate between different classes depending on their total number of citations after $34$ years. We also compare these results between different research fields. In sub-Sec. \ref{MSD}, we study the time averaged MSD, and obtain from it the scaling properties of the auto correlations of the citation number between different years. In sub-Sec.  \ref{Moses}, we quantify the non-stationarity of the citation time series. In sub-Sec. \ref{SecNoah} we study the effect of fat-tails of the yearly citation distribution on the growth of the MSD, and how they change over time. Finally, in sub-Sec. \ref{WeakErgodicity} we compare the non-ergodic behavior in the three fields. The discussion is found in Sec. \ref{SecDiscussion}.  \\

\section{Results} 
\label{SecResults}
\subsection{Citation distribution}
\label{Citationdistribution} 
\begin{figure}[ht!]
\centering
\includegraphics[width=0.5 \textwidth]{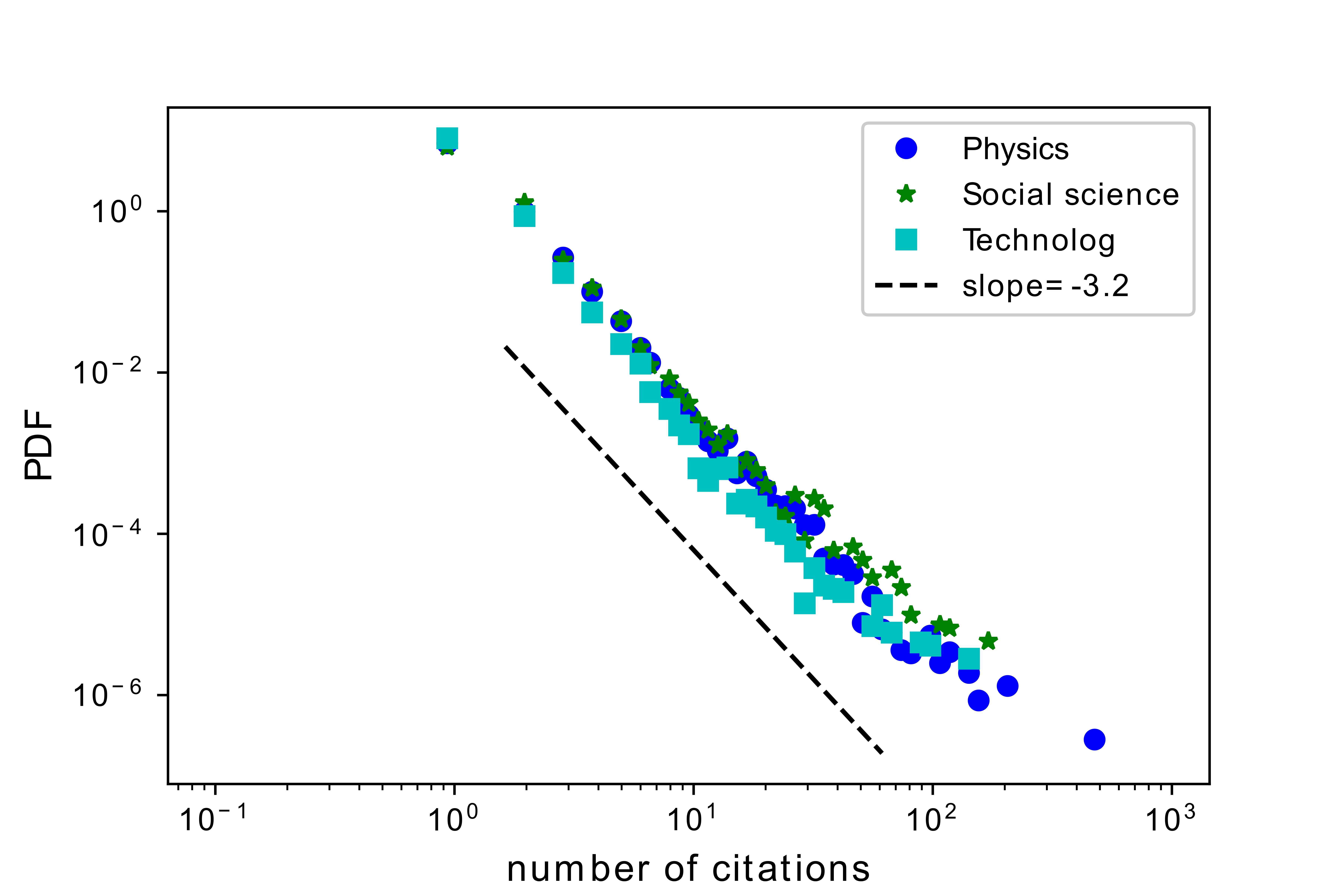}
\caption{\footnotesize{Normalised histograms of citations for three different disciplines at $t=20$ years after publication, show a power-law decay with the exponent of $\approx-3.2$ for all three fields.}}
\label{fig:hist}  
\end{figure}

In this section, we study the normalized distributions of the yearly citation numbers $C_{i}(t)$ of papers published during five years from 1974 to 1978. Table.\ref{tab:1} presents the scientific disciplines and the number of papers whose citation distribution has been analyzed in this paper. 
The length of each series is 34 years, $t=1$ corresponds to the year of the publication. The normalized distributions of $(C_{i}(t))$ in a single year, 20 years after publications, are depicted in figure \ref{fig:hist} for three different disciplines. Figure \ref{fig:hist} shows a power-law decay with a slope of around $-3.2$ for all three distributions, suggesting a qualitatively similar distribution and a universal behavior across these fields.

The power-law nature of the yearly citation distribution for the ISI academic papers and publications, in the journal of Physical Review, is known \cite{clauset2009power,redner1998popular,brzezinski2015power,eom2011characterizing}. However, few other studies reported log-normal distributions as well, and detected universal scaling properties of such distributions in some more specialized categories of science than the ones we are analysing, e.g., nuclear physics and engineering 
%while we considered the whole papers in the field of physics or technology as one category 
\cite{radicchi2008universality,evans2012universality}.   
Note that, probability distributions with power-law tails $\propto 1/C^\gamma$ at large $C$, can be divided into two categories, with significantly different statistical behavior: When $0\leq\gamma\leq2$; the mean-square $\langle C^2\rangle=\int_{-\infty}^\infty C^2P(C)d C\rightarrow\infty$ 
 (when $0\leq\gamma\leq1$, also $\langle |C|\rangle=\int_{-\infty}^\infty |C|P(C)d C\rightarrow\infty$ 
). In this case the dynamics is called ``scale-free``, and no matter at which time-scale we will observe the series $C(t)$, its shape will be governed by only one or few large fluctuations. When $\gamma\geq2$, the first and second moments of $C$ are finite. In our case $\gamma=3.2>2$, as was also concluded by Golosovsky\cite{golosovsky2017power}, the yearly citation distribution does not appear to be scale-free. However in Sec. \ref{SecNoah}, we will revisit this result, adding a second observation, which may point out to the opposite conclusion. 

\begin{table}[t]
\footnotesize
\centering 
\begin{tabular}{ c c c }
\hline\hline 
\hline
 Research field & Years of publication & Number of papers  \\ 
[0.5ex] %
\hline
 Physics & 1974-1978 & 506641 \\  
%%%%% 
 Social science & 1974-1978 & 119704 \\  
%%%%% 
 Technology & 1974-1978  & 368989 \\ 
[0.5ex]
\hline\hline 
\end{tabular}
\caption{\footnotesize{List of scientific fields and the number of papers in each field which their citation time series are analyzed in this paper.}}
\label{tab:1} 
\end{table}

\begin{figure}[ht!]
\centering
\includegraphics[width=0.99 \textwidth]{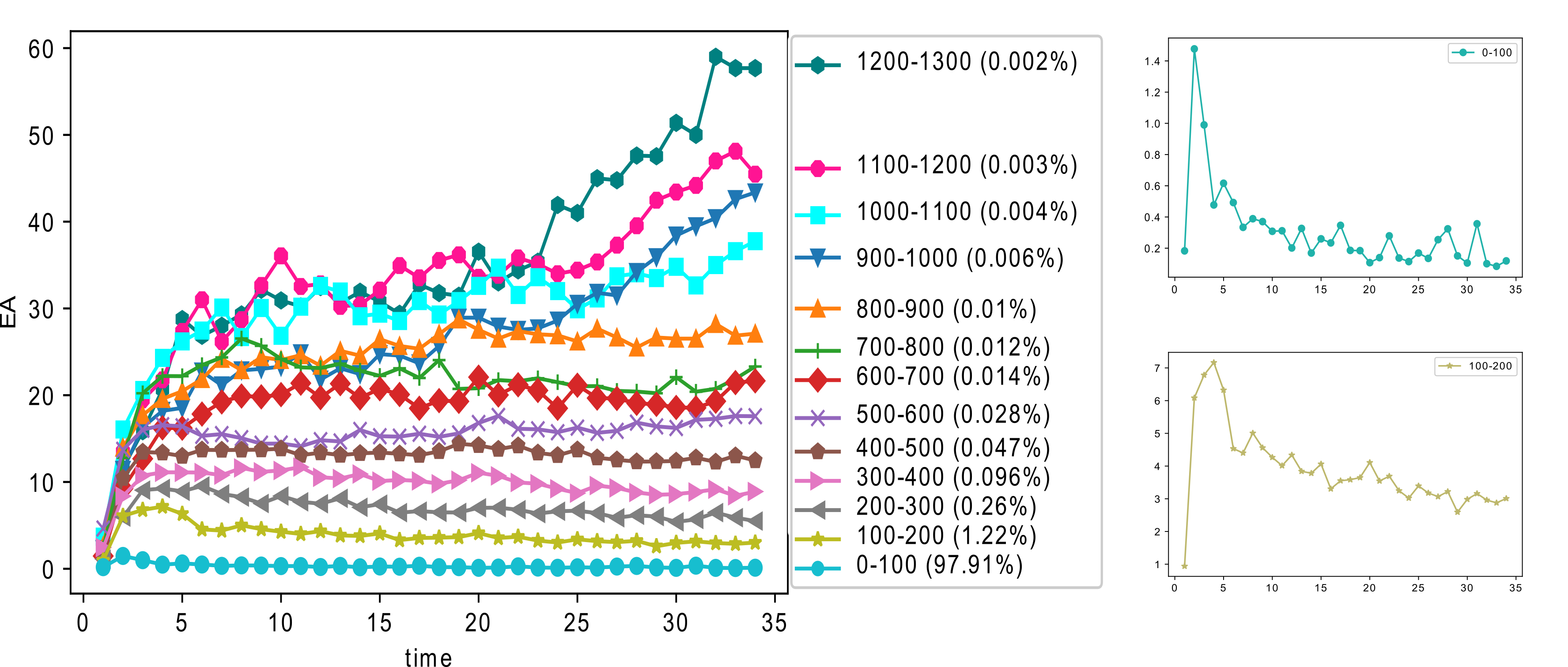}
\caption{ \footnotesize{A comparison of Ensemble Average (EA) for different groups of papers based on the total number of citations in the field of physics after $34$ years. EA of highly cited papers with citations over $900$ continue to grow while EA of papers with number of citations less than this threshold reach to a saturation. The right panel shows the enlargements of the EA versus time for papers with the total citations$\leq 100$ and $100<$citations$\leq 200$.}}
\label{fig:EA_Compare}
\end{figure}

\begin{figure}[ht!]
\centering
\includegraphics[width=0.99 \textwidth]{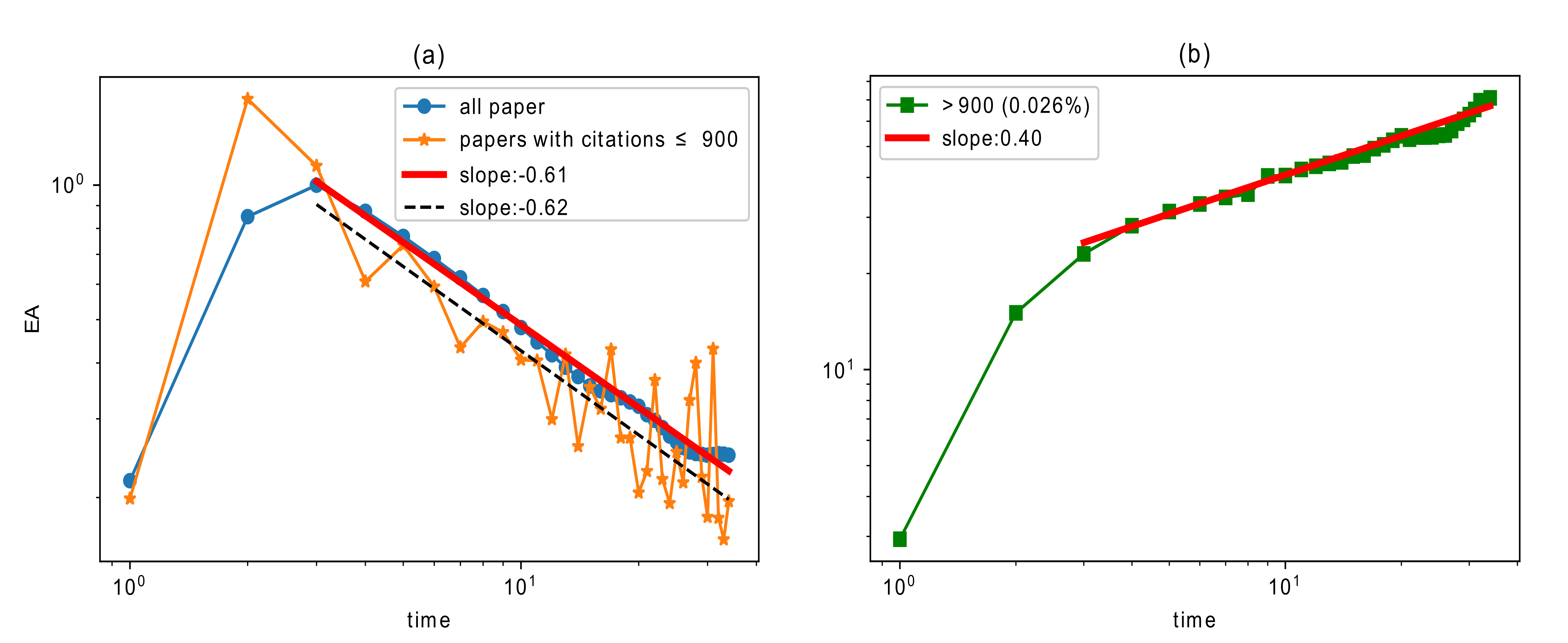}
\caption{\footnotesize{Ensemble Average (EA) of citation time series for physics papers published between 1974-1978 for (a) all the papers (circle) and papers with citations less than 900 (star), showing a power-law decay of the exponent around $-0.61$ for both groups and (b) papers with citations more than $900$, includes only $0.026\%$ of the physics papers, has a power-law growth with exponent $0.4$.}}
\label{fig:EA_Physics_Loglog}
\end{figure}

\begin{figure*}[ht!]
\centering
\includegraphics[width=0.99 \textwidth]{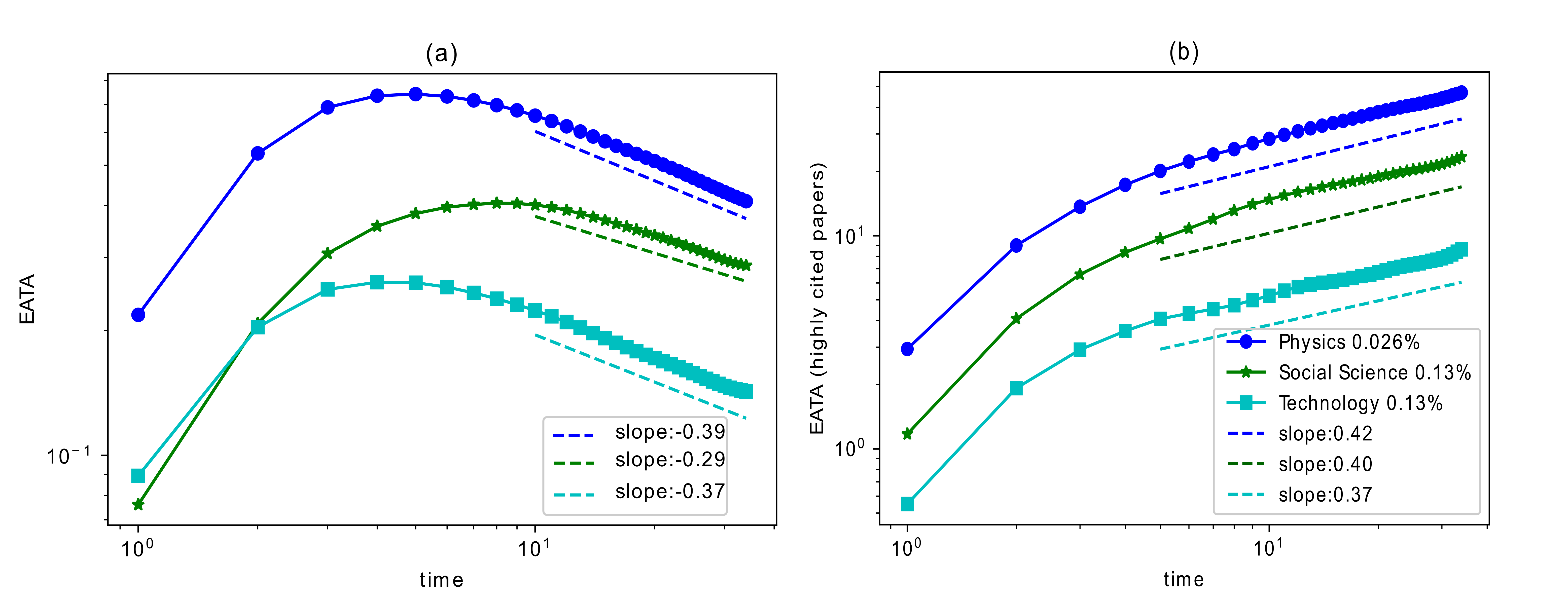}
\caption{\footnotesize{Ensemble Average of Time Average (EATA) for three different fields; physics, social-science and technology. (a) The comparison between all the analyzed data show a power-law decay with almost similar exponents in the three fields. (b) Similarity between highly cited papers in varying fields. Note, the highly cited publications include different percentages of the total number of publications in the different fields which are chosen based on threshold value ($N_{c}$) in the respective field.}}
\label{fig:EATA_diffField}
\end{figure*}

\begin{figure}[ht!]
\centering
\includegraphics[width=0.5 \textwidth]{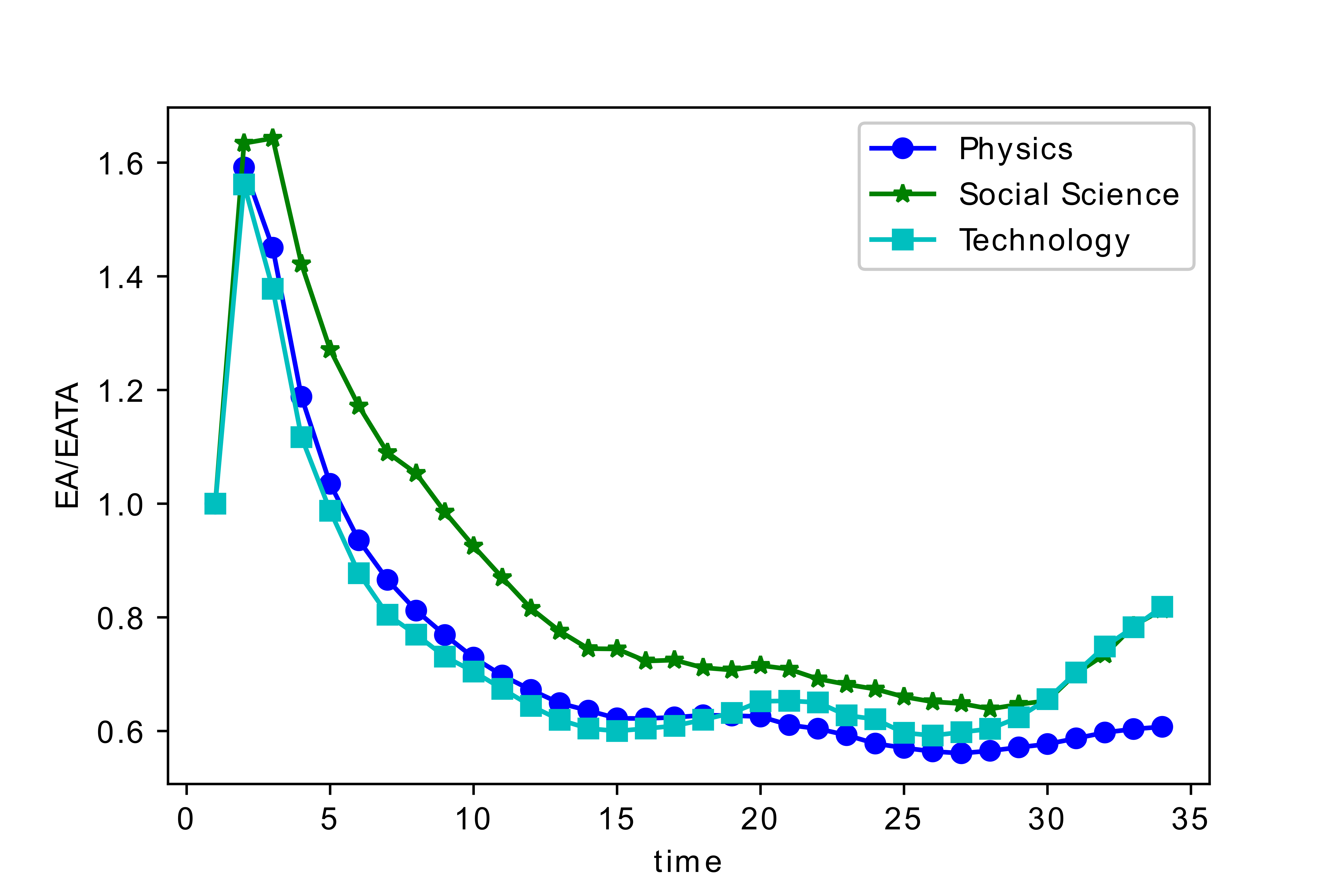}
\caption{\footnotesize{Ensemble Average of Time Average over Ensemble Average of $c_{i}(t)$ saturates at $\approx 0.6$ for the fields of physics and technology and around $\approx 0.7$ for the field of social science.}}
\label{fig:erg_Compare} 
\end{figure}

\begin{figure*}[ht!]
\centering
\includegraphics[width=0.99 \textwidth]{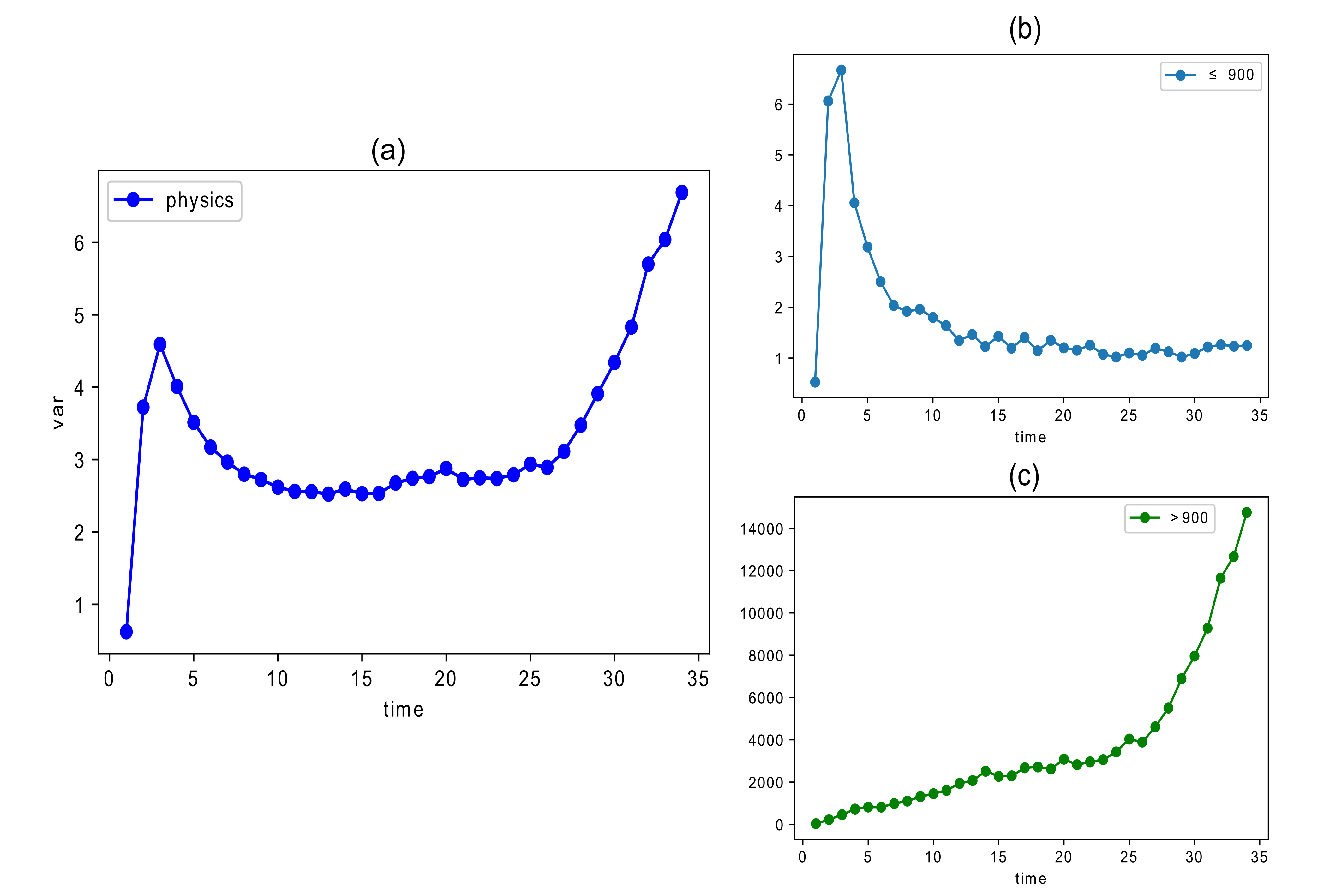}
\caption{\footnotesize{Change of variance of citation time series over time for (a) whole physics papers, (b) papers with less than $900$ citations, (c) highly cited papers (more than 900 citations). For highly cited publications, the variance of citations grows continuously and the speed of growth increases $25$ years after publication.}}
\label{fig:Variance_Phys} 
\end{figure*}

\subsection{The Mean and Variance} 
\label{MeanVariance}
As discussed in the introduction, we study the EA of the citations at each year after publication, up until 34 years, as well as TAs. 
Figure. \ref{fig:EA_Compare} shows the EA of the citations for different groups of papers, for example, in the field of physics, distinguished by the number of citations in each category. The largest category belongs to the papers with citations less than $100$, includes 496085 papers,  which stand for  $97.91\%$ of all the analyzed papers in this field. As shown in this figure, the EA of the lowest categories decreases to zero at long times, whereas in the middle categories,  such as 400-500, at least in our measurement period it seems to reach saturation at a value different from zero. In the categories of papers with more than $N_{c}=900$ citations which, when combined, include only $0.026\%$ of the whole ensemble of analyzed works, the EA continues to grow indefinitely (within our time window), as these famous, highly cited papers obtain increasingly more attention with time. Note that the rate of growth is also changing, after approximately $20$ years from the time of publication. The threshold for the number of citations  beyond which the EA starts to grow continuously, is changing between different fields of science. For instance, $N_{c}=900$ in physics, $N_{c}=400$ for social science and $N_{c}=150$ in technology. These variations show different levels of competition in different fields, and imply for example that ``getting more popular in the field of physics is harder than the other two fields``.\\
In a previous study, Golosovsky \cite{golosovsky2017power} reported that citation time series of the low-cited papers which are published in the same year, reach to a saturation after 10-15 years, while the highly cited papers grow indefinitely. This phenomenon, in which "the rich gets richer and the poor gets poorer", is known in scientific publications \cite{bianconi2001bose}, and is explained as an outcome of the growth rule in complex networks which is called preferential attachment\cite{albert2002statistical,krapivsky2001organization}. 

Figure. \ref{fig:EA_Physics_Loglog}-(a) shows the ensemble average of the citations for all of the analyzed papers in physics (in the other fields of research, our analysis showed qualitatively similar results), as well as papers with citations less than $900$ on logarithmic scale. For the low cited papers and the total publications, we observe a power-law decay with the exponent $\approx-0.61$. The similarity between the two slopes demonstrates the dominance of the low-cited papers on the behavior of the full ensemble because of the sheer number of papers in this group. In figure. \ref{fig:EA_Physics_Loglog}-(b), the EA is shown for the highly cited physics papers (with more than $900$ citations) which shows a power-law growth with the exponent $\approx0.4$. \\
In figure \ref{fig:EATA_diffField}, we present the EATA, and its comparison between three different research fields. As we mentioned above, the threshold value ($N_{c}$) is changing from field to field, which changes the percentage of what we refer to as ``highly cited papers". 
When comparing between EAs and EATAs in figure. \ref{fig:erg_Compare}, where we plot the ratio $\frac{EA}{EATA}$ over time, we can see that in all cases this ratio saturates at some fixed value. This is no surprise, since naturally if the EA is $\propto t^\alpha$, where $\alpha$ is a constant exponent, then clearly the TA $\sim \frac{1}{t}\int EA(t')d t'$ has to have a similar exponent. But since exact values are also important, and not only the qualitative growth of the power-laws, one should note that the \textit{constant limit} to which this ratio converges in this case is $\alpha+1$. In figure \ref{fig:erg_Compare} for physics and technology, it saturates around $\approx0.6$, and for social science,  $\approx0.7$. These values are different from the expected $\alpha+1$, since EA and EATA do not scale linearly in the whole time of study, therefore their corresponding exponents are not equal. \\
Note that it's not only the corresponding EAs of highly cited papers which grow in time, but their variance also grows; $var=\langle C^{2}(t)\rangle-\langle C(t)\rangle^{2}$. Figure. \ref{fig:Variance_Phys}-(a) demonstrates the change of the variance over time for citation time series of the whole physics papers. The variance starts to grow after $\approx25$ years of publications, this behavior is the result of the dominance of highly cited publications on the statistics of the total citations and could be understood  by analysing the citations of low and highly cited papers separately. For low cited publications (figure.\ref{fig:Variance_Phys}-(b)), the variance decreases in time and saturates around $1$. However in figure. \ref{fig:Variance_Phys}-(c), for highly cited ones with citations more than $900$, the variance grows indefinitely and the speed of growth even increases after $25$ years. 

%%%%%%%%%%%%%%%%%%%%%%%%%%%%%%%%%%%%%%%%%%%%%%%%%%%%
\subsection{Time averaged MSD and the correlation} 
\label{MSD}
Let $Y_i(t)=\sum_{n=0}^t C_i(n)$, be the cumulative sum of the yearly citation number, for some paper $i$, until the year $t$ after its publication. Here, we study the fluctuations of the citation trajectories, using the Time Averaged Mean Square Displacement (TA-MSD). For a single trajectory (citation history of one particular paper $i$), TA-MSD is given by, \cite{klafter2011first};
\begin{equation}
   \overline{\delta^{2}(\Delta)}=\frac{1}{T-\Delta}\sum_{t'=0}^{t'=T-\Delta}[Y_{i}(t'+\Delta)-Y_{i}(t')]^{2}. 
   \label{TAMSD}
\end{equation}
The above moving average sums the number of citations added for each trajectory, at intervals of duration  $\Delta$, until time $T-\Delta$, and divided by $T-\Delta$. $t=T$ is the maximal measured time in our data, which is $T=34$ years. The ensemble-averaged TA-MSD is: $\langle \overline{\delta^{2}(\Delta)} \rangle= \frac{1}{N} \sum_{i=1}^{N}\overline{\delta_{i}^{2}(\Delta)}$. 
%\begin{equation}\label{EATAMSD}
%  \langle \overline{\delta^{2}(\Delta)} \rangle= \frac{1}{N} \sum_{i=1}^{N}\overline{\delta_{i}^{2}(\Delta)}.
%\end{equation}
In the data analysis, we consider the maximum lag-time as $\Delta=|\frac{T}{3}|$. Recently, the TA-MSD has been shown to scale as  \cite{meyer2018anomalous,aghion2020moses} 
\begin{equation}
  \langle \overline{\delta^{2}(\Delta)} \rangle \sim t^{2H-2J}\Delta^{2J},
  \label{Joseph}
\end{equation}
where $J\in[0,1]$ is called the {\em Joseph} exponent, which is associated with the autocorrelations in the time series. For a random process without long-ranged autocorrelations; $J=\frac{1}{2}$. 
If a process is long-ranged and positively correlated; $\frac{1}{2}<J \leq 1$, and for an anti-correlated process $0<J<\frac{1}{2}$ \cite{mandelbrot1968noah,lim2002self,chen2017anomalous}. $H$ is called the Hurst exponent, which also quantifies the temporal growth of the MSD; $\langle Y^2\rangle\propto t^{2H}$ at long $t$ \cite{mandelbrot1968noah}. In standard Gaussian processes; $H=1/2$, whereas $H>1/2$ indicates that the process is super-diffusive, and $H<1/2$ is sub-diffusive. 

\begin{figure}[ht!]
\centering
\includegraphics[width=0.99 \textwidth]{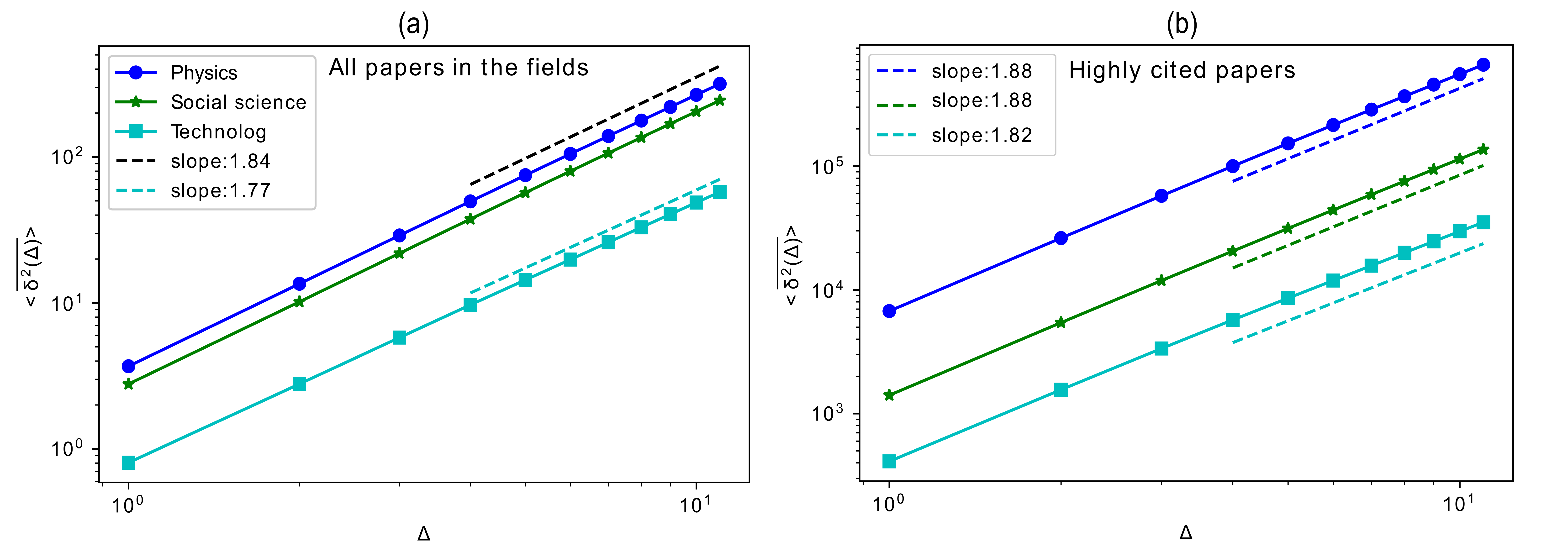}
\caption{\footnotesize{Ensemble average of time average mean square displacement versus lag time $\Delta$, show a scaling behavior with slope $2J$ for (a) whole publications in three fields and (b) highly cited papers in the fields. $J$ is a measure of correlation in time series, show citation trajectories are highly correlated.}}
\label{fig:EATAMSD_Delta} 
\end{figure}

%\begin{figure*}[h!]
%\includegraphics[width=0.99 \textwidth]{MSD_time_loglog.png}
%\caption{\label{fig:MSD_Time_Phys} Ensemble average time average mean square displacement versus time $t$ with constant $\Delta=4$ for (a) highly cited papers with citations more than $900$ and (b) publications with citations less than $900$.}
%\end{figure*}

Figure. \ref{fig:EATAMSD_Delta} displays the scaling of $\langle \overline{\delta^{2}(\Delta)} \rangle$ with respect to $\Delta$ for all the publications in the three mentioned research fields, as well as for the highly cited papers. According to equation \ref{Joseph}, the slopes in figure \ref{fig:EATAMSD_Delta} equals $2J$. Therefore, the Joseph exponent in  all the citation trajectories is higher than $1/2$, representing a high degree of correlation between the number of citations the papers get in subsequent years, across all the fields. This was to be expected, since if a papers is popular and gets high number of citations after its publication, it keeps growing due to its high exposure, while less popular papers, with low number of citations in the first years, will often stay in the same state in the years that follow. What this shows may be somewhat encouraging for some researchers, or discouraging for others, since it means on one hand our ``citation culture" is not completely arbitrary, but on the other hand: if your work does not succeed in becoming popular quickly, your chances to change this situation later are not so great. This trend of behavior is not only for high and low cited papers but also for averaged cited ones too. \\ 
Note that the Joseph exponent for highly cited papers, is a bit larger than that of the averaged and the lower cited ones.

%%%%%%%%%%%%%%%%%%%%%%%%%%%%%%%%%%%%%%%%%%%%%%%%%%%%
\subsection{Characterizing the nonstationarity of citation trajectories, using the Moses exponent} 
\label{Moses}

\begin{figure*}[ht!]
\centering
\includegraphics[width=0.55 \textwidth]{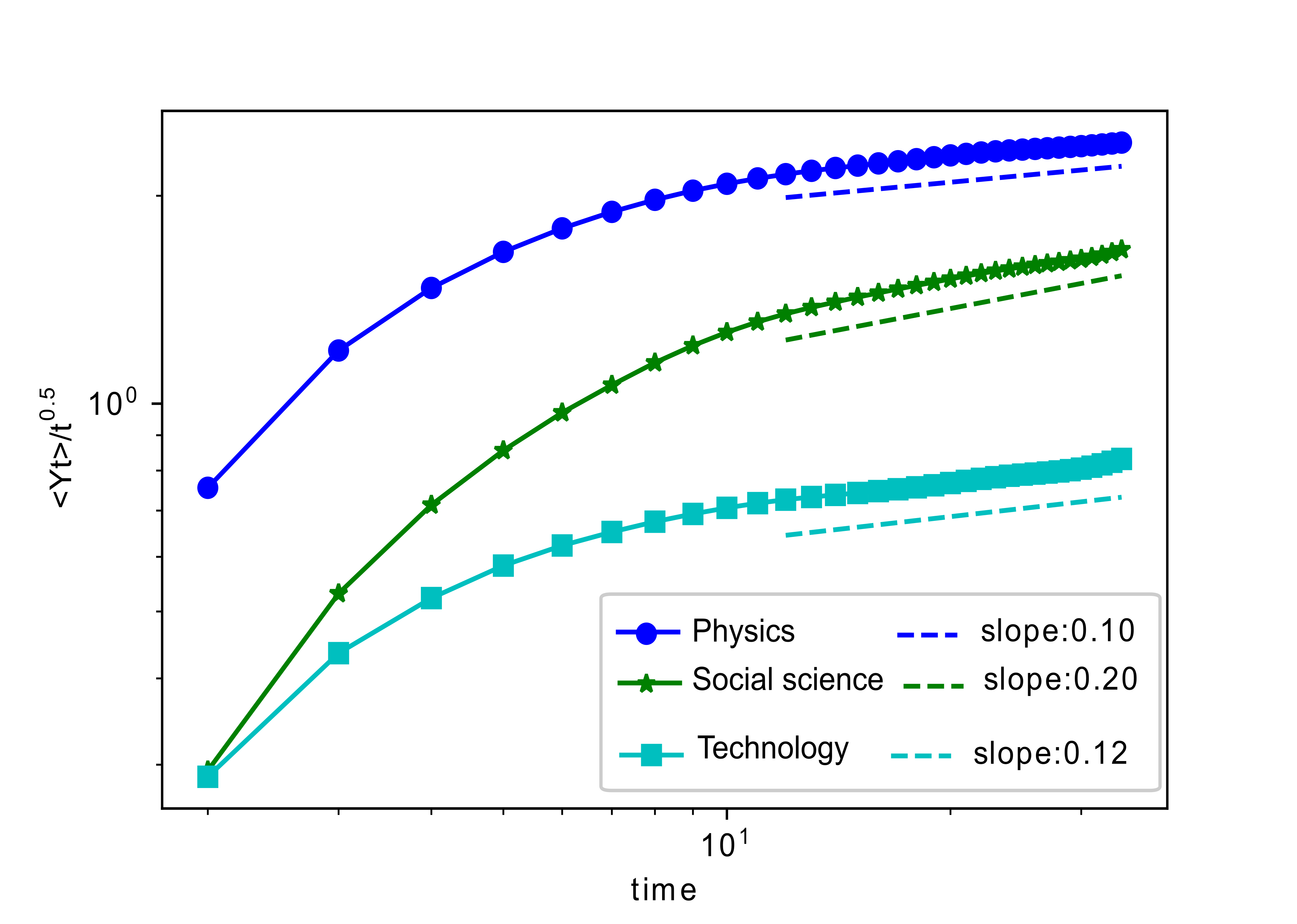}% Here is how to import EPS art
\caption{\footnotesize{Aging process in citation trajectories for all the papers published in 1974-1978 in three different disciplines indicated by a single exponent of $M$. Three fields of science are highly non-stationary with Moses exponent smaller than $0.5$.}}
\label{fig:MosesExp}
\end{figure*}

\begin{figure*}[ht!]
\centering
\includegraphics[width=0.99 \textwidth]{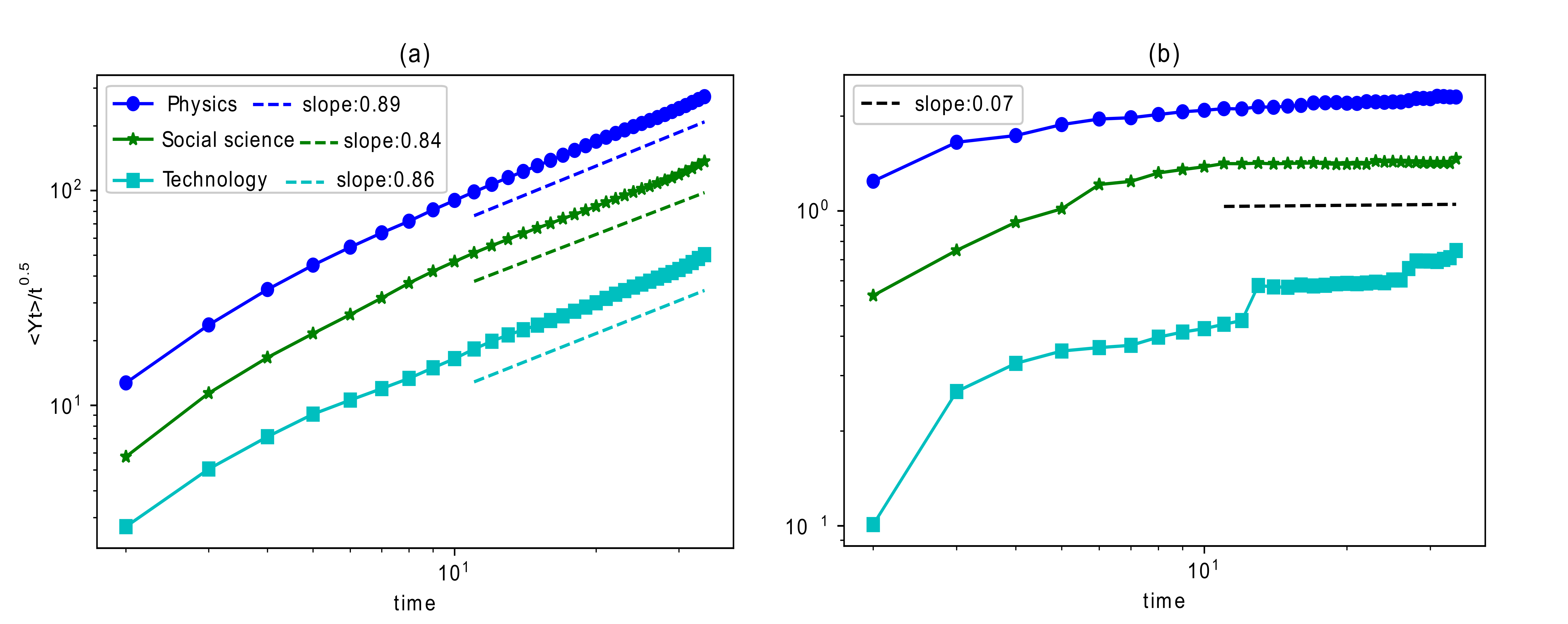}
\caption{\footnotesize{Scaling of the ensemble average of the cumulative sum of the yearly citations versus time for (a) highly cited papers and (b) low cited ones. The values of the Moses exponent for the highly cited papers in all three fields are high, showing the non-aging behavior of these papers in time. However, the low cited papers with Moses exponent close to zero age and die out, in the sense that the rate at which they get new citations only decreases with time}}
\label{fig:MosesExp_Highlow} 
\end{figure*}

In general, in a stochastic process $x_{t}$, if the probability distribution of the increments $(x(t+\tau)-x(t))$ is independent of $t$, the increments are considered stationary. In Refs. \cite{chen2017anomalous,meyer2018anomalous,bormashenko2019moses,aghion2020moses}, it was shown that the level of non-stationarity can be quantified by a single parameter, which is called the Moses exponent. This parameter can be measured directly from the data. Recall that we defined  $Y_i(t)$, as the cumulative sum of the yearly citations of a single paper, after $t$ years.  For a general time-series, the Moses exponent $M$ is defined via the temporal scaling of the EA, of the sum of the absolute-increments of the process \cite{chen2017anomalous}. In our case, since the increments of $Y_i(t)$ are always positive; $|C_{i}(n)|=C_{i}(n)$, we obtain the Moses exponent from  
\begin{equation}
    {\langle Y_{t}\rangle}=t\langle\overline{C_i(t)}\rangle\sim t^{M+\frac{1}{2}},
    \label{MosesExpEq}
\end{equation} 
where, as mentioned above angled brackets mean ensemble average, and overline means time-average. 
For normal diffusive processes, with stationary increments; $M=1/2$. For nonstationary process, when the process dies out since its increments become smaller in time; $M$ is smaller than $1/2$. When we find $M>1/2$, it indicates that the absolute-increments (yearly citations) grow in time. 

Figure. \ref{fig:MosesExp} shows the scaling ${\langle Y_{t}\rangle}/{t^{1/2}}$ versus $t$, for physics, social science and technology. The figure shows that in the first few years the mean of $Y_t$ does not grow like a power-law, as indicated by the non-linear scaling of the log-log plot, but from around $t=10$, the growth is power-law with exponents which are smaller than $1/2$ in all the fields. This demonstrates that overall citation time-series die out in time. To investigate the non-stationarity of this process better, we repeat the analysis for the two groups of low cited and highly cited publications. In figure \ref{fig:MosesExp_Highlow}-(a), the scaling of ${\langle Y_{t}\rangle}/{t^{1/2}}$ for the highly cited papers are shown. $M$ for all three curves are higher than $1/2$, which means that the yearly citations of the most popular papers increase over time and this means that they increase in popularity even regardless of the correlations (recall that the Moses and the Joseph effects are two separate effects \cite{chen2017anomalous}). However, as mentioned above the statistics of highly cited papers have a very small effect on the measured Moses exponent of the whole ensemble, since they are few in number. For low cited papers in figure \ref{fig:MosesExp_Highlow}-(b), ${\langle Y_{t}\rangle}/{t^{1/2}}$, in all three fields, the exponent $M$ is close to $0$, since a paper stops getting cited and its citation time series reaches a plateau with no new citations added anymore.

%%%%%%%%%%%%%%%%%%%%%%%%%%%%%%%%%%%%%%%%%%%%%%%%%%%%
\subsection{The significance of large, rare fluctuation; the Noah effect in citation trajectories} 
\label{SecNoah} 

In Subsec. \ref{Citationdistribution}, we found that the tails of the yearly-citation distribution have a power-law tail which roughly falls-off $\propto 1/C^{3.2}(t)$. Naively, this seems to suggest that the citation time-series is not scale free, namely that its mean and variance are finite. This is also in agreement with the observation reported in Ref. \cite{golosovsky2017power}. However, we would like now to introduce an additional observation, which might alter this conclusion. 
\\ 
 In time-series analysis, when we do not have a large ensemble of data to obtain clear statistics from, the influence of fat-tails in the increment distribution of the process on the anomalous diffusion, can be measured directly from a small number of sufficiently long paths, via the so-called ``Noah effect" \cite{mandelbrot1968noah,chen2017anomalous,meyer2018anomalous}.  After we have obtained the Moses exponent in Subsec. \ref{Moses}, we now obtain the Latent exponent $L$ \cite{mandelbrot1968noah}, via 
 \begin{equation} 
  Z_{t}=\sum_{s=1}^{s=t} [C_{i}(s)-\langle C_i(s)\rangle]^{2},\qquad\mbox{and}\qquad 
 \langle Z_{t}\rangle \sim t^{2L+2M-1}. 
 \label{NoahExponent}
\end{equation} 
Note that, here we use a sampled ensemble average, which is guaranteed to have a finite value at finite times, as opposed to the theoretical value which will be divergent if the increment distribution has scale-free fat tails. One can also use the sampled-median here, instead of the mean. 
 When $M=1/2$, namely when we do not observe aging effects in the time-series, if $L=1/2$; $ \langle Z_{t}\rangle \sim t$, which is similar to a standard Gaussian process with a finite increment-variance. On the other hand, when $L>1/2$, this can only occur because the increment distribution at least has a regime, with a power-law shape that falls-off as $1/C^\gamma$, and $0<\gamma<2$ \cite{chen2017anomalous,aghion2020moses}. Namely in this regime the distribution does not have a typical value. This power-law shape of the tails of the distribution might have a time-dependent cutoff, which is pushed towards $\infty$ as time increases, but it needs to be visible at a certain regime of $C$ also in finite, but long times. Note that, at least at long times, $L$ can not be smaller than half, since
this would mean that the average of the squared-increments expands more
slowly in time than the square of their mean absolute value. All this
explanation holds also when $M\neq1/2$. 

\begin{figure*}[ht!]
\centering
\includegraphics[width=0.99 \textwidth]{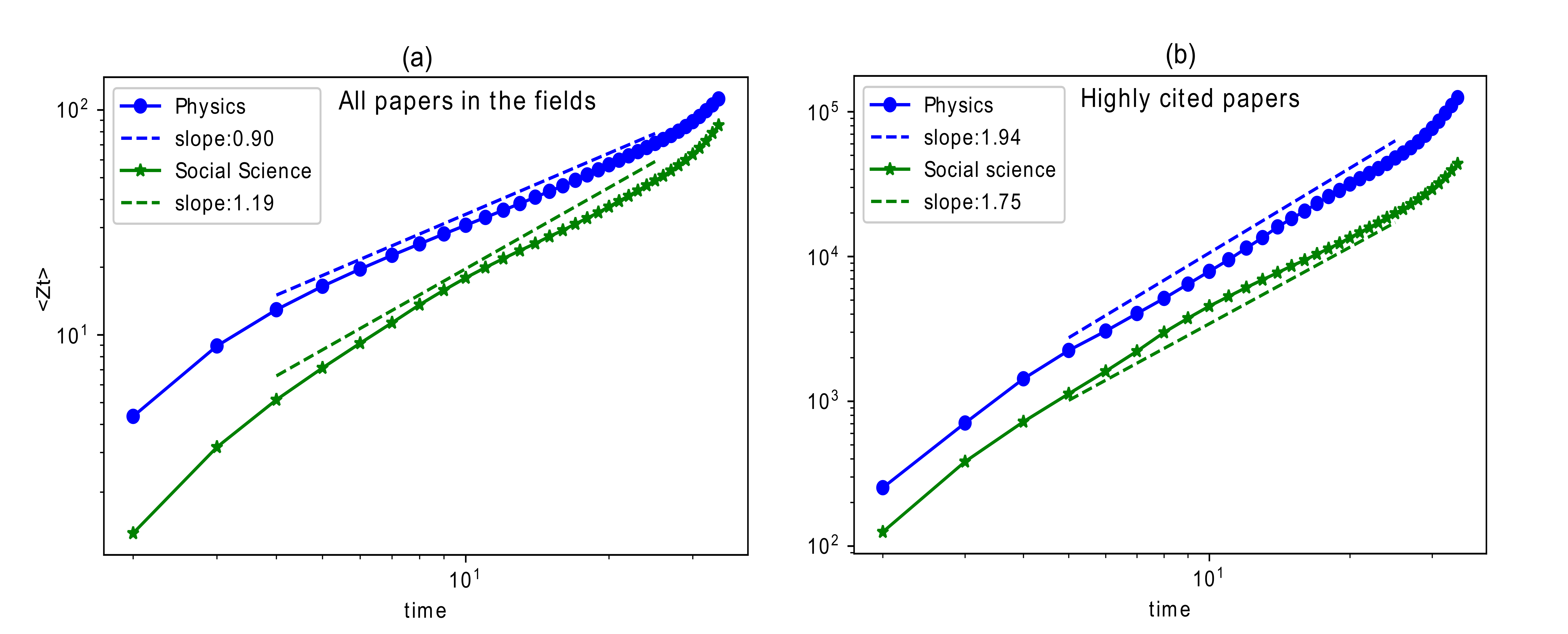}
\caption{\footnotesize{Scaling of $\langle Z_{t} \rangle$ in time for (a) total publications of two fields of physics and social science and (b) highly cited papers in these fields. The result from the field of technology has not been shown here since we did not find a linear scaling behavior for that. Linear fit is done in the region with the best linear behavior, there is a cross-over after $25$ years which is related to the increasing of the speed of growth for variance of yearly citations has been shown in figure \ref{fig:Variance_Phys}.}}
\label{fig:Noah_Exponent} 
\end{figure*}

In figure \ref{fig:Noah_Exponent}, we show the scaling of $\langle Z_{t}\rangle$ for publications of two fields of science (physics and social science) and highly cited papers.  
Figure \ref{fig:Noah_Exponent}-(a) shows the presence of a strong Noah effect, when we observe the full ensemble of trajectories together. This results from the highest cited papers, which obtain far more citations than the others. This Noah effect suggests that the citation data is indeed scale-free, but we could not detect that by observing the shape of the tails of the distribution, since the highly cited papers are too few in number. If we look again at Fig. \ref{Citationdistribution}, we see that clearly beyond the region where we find a linear behavior (in the log-log plot), with a slope $>3$, we find another regime where the slope is $1$. The latter however is only the effect of the logarithmically-sized bins that we used to plot the distribution, and it indicates that in every bin in the far tails, we had found only 1 paper, since the papers are very sparsely distributed. All it means, is that we probably do not have enough data to sample the distribution well enough in the far-most tails, but we suspect that the shape of the tails there indeed corresponds to a  scale-free distribution, whose trace is detected by the Noah effect.\\
Figure \ref{fig:Noah_Exponent}-(b) shows the linear scaling of $\langle Z_{t}\rangle$ for highly cited papers. The Latent exponent, $L$, for these publications is close to $0.5$. Here, the explanation is clear: since we only choose to look at the distribution of a small percentage of all papers, which are located in the same region of the histogram, their own distribution is thin (this sample of the papers is more homogeneous than the total ensemble of all of them). 

%%%%%%%%%%%%%%%%%%%%%%%%%%%%%%%%%%%%%%%%%%%%%%%%%%%% 
\subsection{The Hurst exponent}
\begin{figure*}[ht!]
\centering
\includegraphics[width=0.99 \textwidth]{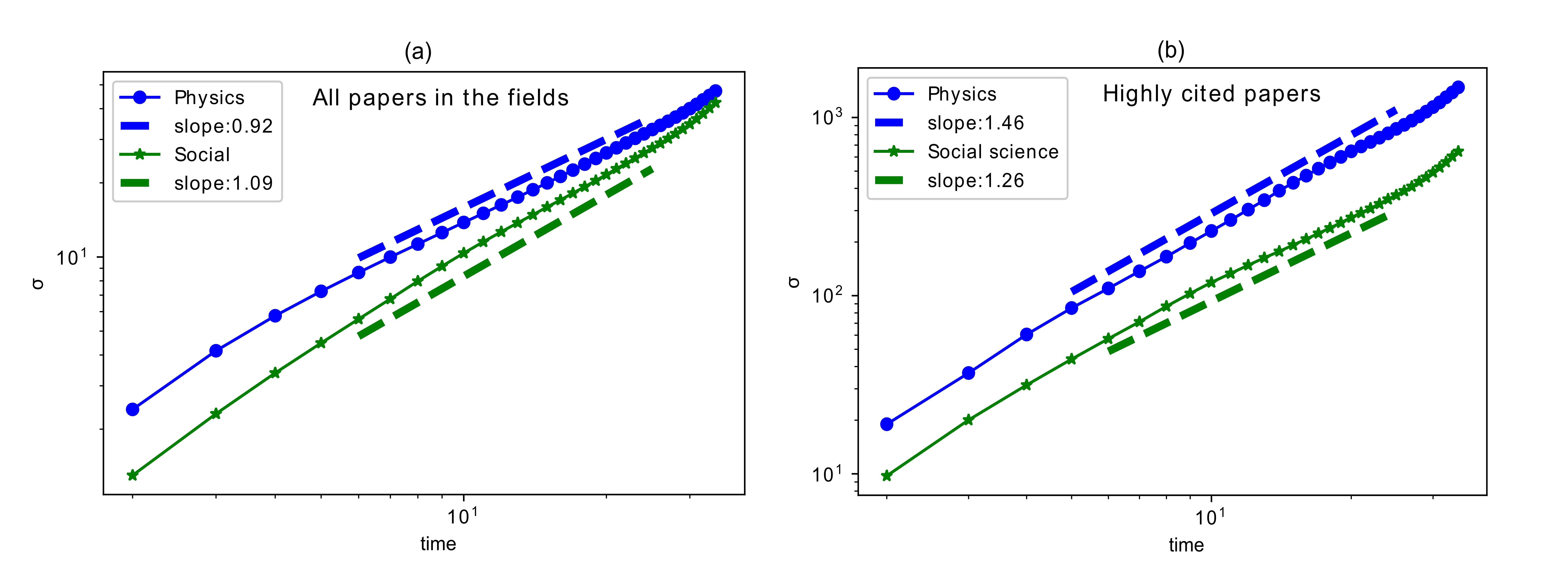}
\caption{\footnotesize{Scaling of the standard-deviation ($\sigma$) of distributions of $Y_{t}$ in time represents the Hurst exponents in the fields of physics and social science for (a) all the publications in the fields. (b) Highly cited papers, the high value of $H$ for these papers shows the super diffusive behaviors of the citation trajectories of these papers. The linear fit has been used in the same range as in figure \ref{fig:Noah_Exponent}.}}
\label{fig:Hurst_Exponent} 
\end{figure*}

So far, we have shown that the yearly citation time series exhibit both correlation, non-stationarity and possibly large fluctuations due stemming from a fat-tailed distribution.  In Refs. \cite{chen2017anomalous,mandelbrot2002gaussian,meyer2018anomalous,aghion2020moses}, it was shown that using the three exponents; $M,L$ and $J$, we can obtain the Hurst exponent 

\begin{equation}
  \sigma^{2}=(\langle Y^2_{t}\rangle - \langle Y_{t}\rangle ^{2}) \sim t^{2H}, 
    \label{Hurst1}  
\end{equation}
via a simple summation relation 
\begin{equation}
  H=J+L+M-1.
  \label{Hurst2}
\end{equation} 
Figure.\ref{fig:Hurst_Exponent}-(a) show the scaling of $\sigma$ with time, for the full ensembles of physics papers and social science, with their corresponding Hurst exponents. Figure. \ref{fig:Hurst_Exponent}-(b) shows the scaling of $\sigma$ with time, obtained only from the highly cited papers. In all the categories clearly the citation data is super-diffusive, and for the highly-cited ones, even super-ballistic. 
In table. \ref{tab:2}, the four exponents for all the publications in the fields as well as for highly cited papers are represented. The relation in equation. \ref{Hurst2} holds nicely in all the different categories. This indicates that one may use this summation relation to obtain  any of the four exponents; $M,L,J,$ and $H$, from the other three, since the correlations/ the non-stationarity and the large fluctuations in the time-series are not decoupled from each-other. 

\begin{table}[t]
\footnotesize
\centering 
\begin{tabular}{ c c c c c c}
\hline\hline
\hline
 Research fields & M & J & L & H & J+L+M-1  \\ 
[0.5ex] %
\hline
 Physics & 0.10 & 0.92 & 0.85 & 0.92 & 0.87  \\  
%%%%%
 Social science & 0.206 & 0.92 & 0.88 & 1.09 & 1.01 \\  
%%%%%
Physics (Highly cited papers) & 0.89 & 0.94 & 0.58 & 1.46 & 1.41  \\  
%%%%%
 Social science (Highly cited papers) & 0.84 & 0.94 & 0.53 & 1.26 & 1.31 \\ [0.5ex]
\hline\hline 
\end{tabular}
\caption{\footnotesize{Research fields of physics and social science and their corresponding exponents.}}
\label{tab:2} 
\end{table}

%%%%%%%%%%%%%%%%%%%%%%%%%%%%%%%%%%%%%%%%%%%%%%%%%%%% 
\subsection{Weak Ergodicity Breaking}
\label{WeakErgodicity} 

According to what we have discussed up to now, the citation time series are highly non-stationary, correlated, have large fluctuations, and consequently they are non-ergodic. In this sub-section, we quantify the weak ergodicity breaking , by measuring the ratio between the ensemble-time average
$\langle \overline{\delta^{2}(\Delta)} \rangle= \frac{1}{N} \sum_{i=1}^{N}\overline{\delta_{i}^{2}(\Delta)}$, where ${\delta_{i}^{2}(\Delta)}$ is defined in Eq. \eqref{TAMSD}, and the ensemble average $\langle (Y_{i} (T+\Delta)-Y_{i}(T))^2\rangle$, as function of the increment duration $\Delta$. 

\begin{figure*}[ht!]
\centering
\includegraphics[width=0.5 \textwidth]{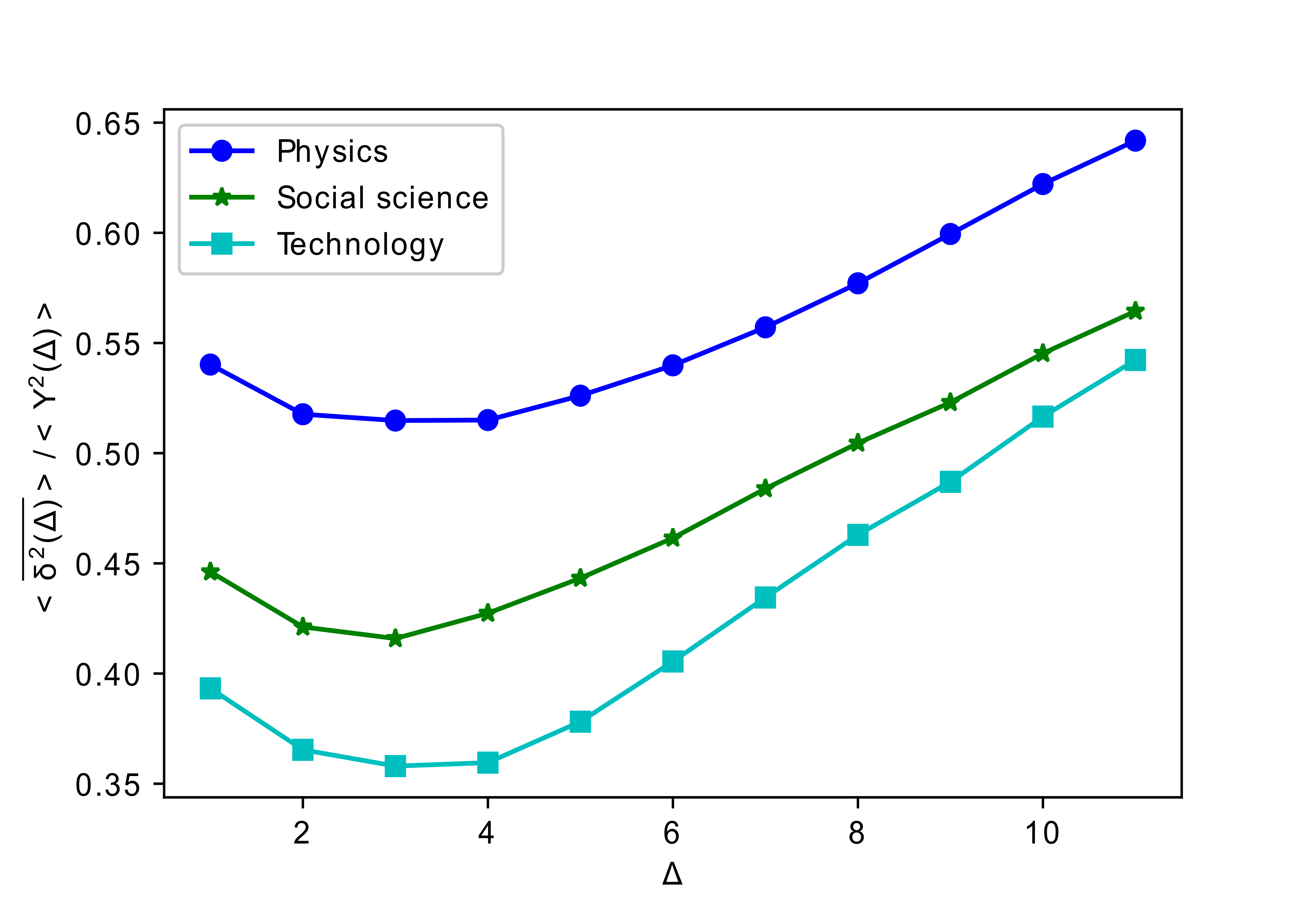}
\caption{\footnotesize{The growth of the ratio $\frac{\langle \overline{\delta^2(\Delta) \rangle}}{\langle Y^{2}(\Delta) \rangle}$, for papers in physics, social science and technology, as function of $\Delta$.}}
\label{fig:Erg}
\end{figure*}

For diffusive processes the ensemble and time average of MSD are not equivalent, $\langle Y^{2}(\Delta) \rangle \neq \overline{\delta^2(\Delta)}$ \cite{bouchaud1992weak}. Figure \ref{fig:Erg} shows that in the three fields of science that we study, this ratio $\frac{\langle \overline{\delta^2(\Delta) \rangle}}{\langle Y^{2}(\Delta) \rangle}$  grows with $\Delta$, so care should be taken when one compares the results of one type of measurement procedure, and the other.

%%%%%%%%%%%%%%%%%%%%%%%%%%
%%%%%%%%%%%%%%%%%%%%%%%%%%
\section{Discussion} 
\label{SecDiscussion}
In this paper, the aging process in the time series of citations to scientific papers has been investigated, by considering first, the total ensemble of publications, and then by separating highly cited papers from less popular ones.  
We found that the anomalous diffusion in the distributions of citations is a results of the non-stationarity of the yearly citation distribution, as well as large fluctuations and temporal correlation. Here we see all the three effects leading to anomalous diffusion, combined. 
\begin{itemize}

    \item Citation time series for the three analyzed fields are highly correlated. The correlation is measured using ensemble average of mean squared displacement and quantified by an exponent called Joseph exponent.  
    \item Citations trajectories are highly non-stationary, this effect has been quantified with a well defined Moses exponent ($M$), for a Gaussian stationary process $M$ equals $0.5$. When $M$ deviates from $0.5$ towards higher values it signifies a growth in the process (papers get more attention in time).  This effect has been observed in popular papers with citations more than $900$, $400$ and $150$ respectively in physics, social science and technology. In a process which ages and dies out in time, the Moses exponent $M$
    is smaller than $0.5 $. This property has been detected for total publications in the fields because of high number of low-cited papers and their dominance in the statistics.  For a low cited publication $M$ is close to zero. 
    \item Citation trajectories show a strong Noah effect with a Latent exponent above $0.8$. This effect suggests a fat-tailed distribution of yearly citations which has not been observed in the tail of distribution because of the sparse number of highly cited papers.

\end{itemize}
All the three factors above  lead to anomalous diffusion of the citation trajectories. The relation between the three scaling exponents and the Hurst exponent holds perfectly for the total publications as well as highly cited ones.  
For the possible future research, it will be interesting to know how the presence of online archives such as Google Scholar as well as new ways of giving exposure to the papers in future might affect the behavior of citation time series.

%\textbf{Acknowledgement}
%\\
%This research is supported by the Max-Planck society.

\bibliographystyle{aipnum4-1}
\bibliography{./bibliography2}

\begin{thebibliography}{42}%
\makeatletter
\providecommand \@ifxundefined [1]{%
 \@ifx{#1\undefined}
}%
\providecommand \@ifnum [1]{%
 \ifnum #1\expandafter \@firstoftwo
 \else \expandafter \@secondoftwo
 \fi
}%
\providecommand \@ifx [1]{%
 \ifx #1\expandafter \@firstoftwo
 \else \expandafter \@secondoftwo
 \fi
}%
\providecommand \natexlab [1]{#1}%
\providecommand \enquote  [1]{``#1''}%
\providecommand \bibnamefont  [1]{#1}%
\providecommand \bibfnamefont [1]{#1}%
\providecommand \citenamefont [1]{#1}%
\providecommand \href@noop [0]{\@secondoftwo}%
\providecommand \href [0]{\begingroup \@sanitize@url \@href}%
\providecommand \@href[1]{\@@startlink{#1}\@@href}%
\providecommand \@@href[1]{\endgroup#1\@@endlink}%
\providecommand \@sanitize@url [0]{\catcode `\\12\catcode `\$12\catcode
  `\&12\catcode `\#12\catcode `\^12\catcode `\_12\catcode `\%12\relax}%
\providecommand \@@startlink[1]{}%
\providecommand \@@endlink[0]{}%
\providecommand \url  [0]{\begingroup\@sanitize@url \@url }%
\providecommand \@url [1]{\endgroup\@href {#1}{\urlprefix }}%
\providecommand \urlprefix  [0]{URL }%
\providecommand \Eprint [0]{\href }%
\providecommand \doibase [0]{http://dx.doi.org/}%
\providecommand \selectlanguage [0]{\@gobble}%
\providecommand \bibinfo  [0]{\@secondoftwo}%
\providecommand \bibfield  [0]{\@secondoftwo}%
\providecommand \translation [1]{[#1]}%
\providecommand \BibitemOpen [0]{}%
\providecommand \bibitemStop [0]{}%
\providecommand \bibitemNoStop [0]{.\EOS\space}%
\providecommand \EOS [0]{\spacefactor3000\relax}%
\providecommand \BibitemShut  [1]{\csname bibitem#1\endcsname}%
\let\auto@bib@innerbib\@empty
%</preamble>
\bibitem [{\citenamefont {Kuhn}(2012)}]{kuhn2012structure}%
  \BibitemOpen
  \bibfield  {author} {\bibinfo {author} {\bibfnamefont {T.~S.}\ \bibnamefont
  {Kuhn}},\ }\href@noop {} {\emph {\bibinfo {title} {The structure of
  scientific revolutions}}}\ (\bibinfo  {publisher} {University of Chicago
  press},\ \bibinfo {year} {2012})\BibitemShut {NoStop}%
\bibitem [{\citenamefont {Evans}(2013)}]{evans2013future}%
  \BibitemOpen
  \bibfield  {author} {\bibinfo {author} {\bibfnamefont {J.~A.}\ \bibnamefont
  {Evans}},\ }\href@noop {} {\bibfield  {journal} {\bibinfo  {journal}
  {science}\ }\textbf {\bibinfo {volume} {342}},\ \bibinfo {pages} {44}
  (\bibinfo {year} {2013})}\BibitemShut {NoStop}%
\bibitem [{\citenamefont {King}\ \emph {et~al.}(2009)\citenamefont {King},
  \citenamefont {Rowland}, \citenamefont {Oliver}, \citenamefont {Young},
  \citenamefont {Aubrey}, \citenamefont {Byrne}, \citenamefont {Liakata},
  \citenamefont {Markham}, \citenamefont {Pir}, \citenamefont {Soldatova} \emph
  {et~al.}}]{king2009automation}%
  \BibitemOpen
  \bibfield  {author} {\bibinfo {author} {\bibfnamefont {R.~D.}\ \bibnamefont
  {King}}, \bibinfo {author} {\bibfnamefont {J.}~\bibnamefont {Rowland}},
  \bibinfo {author} {\bibfnamefont {S.~G.}\ \bibnamefont {Oliver}}, \bibinfo
  {author} {\bibfnamefont {M.}~\bibnamefont {Young}}, \bibinfo {author}
  {\bibfnamefont {W.}~\bibnamefont {Aubrey}}, \bibinfo {author} {\bibfnamefont
  {E.}~\bibnamefont {Byrne}}, \bibinfo {author} {\bibfnamefont
  {M.}~\bibnamefont {Liakata}}, \bibinfo {author} {\bibfnamefont
  {M.}~\bibnamefont {Markham}}, \bibinfo {author} {\bibfnamefont
  {P.}~\bibnamefont {Pir}}, \bibinfo {author} {\bibfnamefont {L.~N.}\
  \bibnamefont {Soldatova}},  \emph {et~al.},\ }\href@noop {} {\bibfield
  {journal} {\bibinfo  {journal} {Science}\ }\textbf {\bibinfo {volume}
  {324}},\ \bibinfo {pages} {85} (\bibinfo {year} {2009})}\BibitemShut
  {NoStop}%
\bibitem [{\citenamefont {Wu}, \citenamefont {Wang},\ and\ \citenamefont
  {Evans}(2019)}]{wu2019large}%
  \BibitemOpen
  \bibfield  {author} {\bibinfo {author} {\bibfnamefont {L.}~\bibnamefont
  {Wu}}, \bibinfo {author} {\bibfnamefont {D.}~\bibnamefont {Wang}}, \ and\
  \bibinfo {author} {\bibfnamefont {J.~A.}\ \bibnamefont {Evans}},\ }\href@noop
  {} {\bibfield  {journal} {\bibinfo  {journal} {Nature}\ }\textbf {\bibinfo
  {volume} {566}},\ \bibinfo {pages} {378} (\bibinfo {year}
  {2019})}\BibitemShut {NoStop}%
\bibitem [{\citenamefont {Deville}\ \emph {et~al.}(2014)\citenamefont
  {Deville}, \citenamefont {Wang}, \citenamefont {Sinatra}, \citenamefont
  {Song}, \citenamefont {Blondel},\ and\ \citenamefont
  {Barab{\'a}si}}]{deville2014career}%
  \BibitemOpen
  \bibfield  {author} {\bibinfo {author} {\bibfnamefont {P.}~\bibnamefont
  {Deville}}, \bibinfo {author} {\bibfnamefont {D.}~\bibnamefont {Wang}},
  \bibinfo {author} {\bibfnamefont {R.}~\bibnamefont {Sinatra}}, \bibinfo
  {author} {\bibfnamefont {C.}~\bibnamefont {Song}}, \bibinfo {author}
  {\bibfnamefont {V.~D.}\ \bibnamefont {Blondel}}, \ and\ \bibinfo {author}
  {\bibfnamefont {A.-L.}\ \bibnamefont {Barab{\'a}si}},\ }\href@noop {}
  {\bibfield  {journal} {\bibinfo  {journal} {Scientific reports}\ }\textbf
  {\bibinfo {volume} {4}},\ \bibinfo {pages} {1} (\bibinfo {year}
  {2014})}\BibitemShut {NoStop}%
\bibitem [{\citenamefont {Sinatra}\ \emph {et~al.}(2016)\citenamefont
  {Sinatra}, \citenamefont {Wang}, \citenamefont {Deville}, \citenamefont
  {Song},\ and\ \citenamefont {Barab{\'a}si}}]{sinatra2016quantifying}%
  \BibitemOpen
  \bibfield  {author} {\bibinfo {author} {\bibfnamefont {R.}~\bibnamefont
  {Sinatra}}, \bibinfo {author} {\bibfnamefont {D.}~\bibnamefont {Wang}},
  \bibinfo {author} {\bibfnamefont {P.}~\bibnamefont {Deville}}, \bibinfo
  {author} {\bibfnamefont {C.}~\bibnamefont {Song}}, \ and\ \bibinfo {author}
  {\bibfnamefont {A.-L.}\ \bibnamefont {Barab{\'a}si}},\ }\href@noop {}
  {\bibfield  {journal} {\bibinfo  {journal} {Science}\ }\textbf {\bibinfo
  {volume} {354}} (\bibinfo {year} {2016})}\BibitemShut {NoStop}%
\bibitem [{\citenamefont {Sekara}\ \emph {et~al.}(2018)\citenamefont {Sekara},
  \citenamefont {Deville}, \citenamefont {Ahnert}, \citenamefont
  {Barab{\'a}si}, \citenamefont {Sinatra},\ and\ \citenamefont
  {Lehmann}}]{sekara2018chaperone}%
  \BibitemOpen
  \bibfield  {author} {\bibinfo {author} {\bibfnamefont {V.}~\bibnamefont
  {Sekara}}, \bibinfo {author} {\bibfnamefont {P.}~\bibnamefont {Deville}},
  \bibinfo {author} {\bibfnamefont {S.~E.}\ \bibnamefont {Ahnert}}, \bibinfo
  {author} {\bibfnamefont {A.-L.}\ \bibnamefont {Barab{\'a}si}}, \bibinfo
  {author} {\bibfnamefont {R.}~\bibnamefont {Sinatra}}, \ and\ \bibinfo
  {author} {\bibfnamefont {S.}~\bibnamefont {Lehmann}},\ }\href@noop {}
  {\bibfield  {journal} {\bibinfo  {journal} {Proceedings of the National
  Academy of Sciences}\ }\textbf {\bibinfo {volume} {115}},\ \bibinfo {pages}
  {12603} (\bibinfo {year} {2018})}\BibitemShut {NoStop}%
\bibitem [{\citenamefont {Fortunato}\ \emph {et~al.}(2018)\citenamefont
  {Fortunato}, \citenamefont {Bergstrom}, \citenamefont {B{\"o}rner},
  \citenamefont {Evans}, \citenamefont {Helbing}, \citenamefont
  {Milojevi{\'c}}, \citenamefont {Petersen}, \citenamefont {Radicchi},
  \citenamefont {Sinatra}, \citenamefont {Uzzi} \emph
  {et~al.}}]{fortunato2018science}%
  \BibitemOpen
  \bibfield  {author} {\bibinfo {author} {\bibfnamefont {S.}~\bibnamefont
  {Fortunato}}, \bibinfo {author} {\bibfnamefont {C.~T.}\ \bibnamefont
  {Bergstrom}}, \bibinfo {author} {\bibfnamefont {K.}~\bibnamefont
  {B{\"o}rner}}, \bibinfo {author} {\bibfnamefont {J.~A.}\ \bibnamefont
  {Evans}}, \bibinfo {author} {\bibfnamefont {D.}~\bibnamefont {Helbing}},
  \bibinfo {author} {\bibfnamefont {S.}~\bibnamefont {Milojevi{\'c}}}, \bibinfo
  {author} {\bibfnamefont {A.~M.}\ \bibnamefont {Petersen}}, \bibinfo {author}
  {\bibfnamefont {F.}~\bibnamefont {Radicchi}}, \bibinfo {author}
  {\bibfnamefont {R.}~\bibnamefont {Sinatra}}, \bibinfo {author} {\bibfnamefont
  {B.}~\bibnamefont {Uzzi}},  \emph {et~al.},\ }\href@noop {} {\bibfield
  {journal} {\bibinfo  {journal} {Science}\ }\textbf {\bibinfo {volume} {359}}
  (\bibinfo {year} {2018})}\BibitemShut {NoStop}%
\bibitem [{\citenamefont {Zamani}\ \emph {et~al.}(2020)\citenamefont {Zamani},
  \citenamefont {Tejedor}, \citenamefont {Vogl}, \citenamefont {Kr{\"a}utli},
  \citenamefont {Valleriani},\ and\ \citenamefont
  {Kantz}}]{zamani2020evolution}%
  \BibitemOpen
  \bibfield  {author} {\bibinfo {author} {\bibfnamefont {M.}~\bibnamefont
  {Zamani}}, \bibinfo {author} {\bibfnamefont {A.}~\bibnamefont {Tejedor}},
  \bibinfo {author} {\bibfnamefont {M.}~\bibnamefont {Vogl}}, \bibinfo {author}
  {\bibfnamefont {F.}~\bibnamefont {Kr{\"a}utli}}, \bibinfo {author}
  {\bibfnamefont {M.}~\bibnamefont {Valleriani}}, \ and\ \bibinfo {author}
  {\bibfnamefont {H.}~\bibnamefont {Kantz}},\ }\href@noop {} {\bibfield
  {journal} {\bibinfo  {journal} {Scientific Reports}\ }\textbf {\bibinfo
  {volume} {10}},\ \bibinfo {pages} {1} (\bibinfo {year} {2020})}\BibitemShut
  {NoStop}%
\bibitem [{\citenamefont {Garfield}(1955)}]{garfield1955citation}%
  \BibitemOpen
  \bibfield  {author} {\bibinfo {author} {\bibfnamefont {E.}~\bibnamefont
  {Garfield}},\ }\href@noop {} {\bibfield  {journal} {\bibinfo  {journal}
  {Science}\ }\textbf {\bibinfo {volume} {122}},\ \bibinfo {pages} {108}
  (\bibinfo {year} {1955})}\BibitemShut {NoStop}%
\bibitem [{\citenamefont {Radicchi}, \citenamefont {Fortunato},\ and\
  \citenamefont {Castellano}(2008)}]{radicchi2008universality}%
  \BibitemOpen
  \bibfield  {author} {\bibinfo {author} {\bibfnamefont {F.}~\bibnamefont
  {Radicchi}}, \bibinfo {author} {\bibfnamefont {S.}~\bibnamefont {Fortunato}},
  \ and\ \bibinfo {author} {\bibfnamefont {C.}~\bibnamefont {Castellano}},\
  }\href@noop {} {\bibfield  {journal} {\bibinfo  {journal} {Proceedings of the
  National Academy of Sciences}\ }\textbf {\bibinfo {volume} {105}},\ \bibinfo
  {pages} {17268} (\bibinfo {year} {2008})}\BibitemShut {NoStop}%
\bibitem [{\citenamefont {Wang}, \citenamefont {Song},\ and\ \citenamefont
  {Barab{\'a}si}(2013)}]{wang2013quantifying}%
  \BibitemOpen
  \bibfield  {author} {\bibinfo {author} {\bibfnamefont {D.}~\bibnamefont
  {Wang}}, \bibinfo {author} {\bibfnamefont {C.}~\bibnamefont {Song}}, \ and\
  \bibinfo {author} {\bibfnamefont {A.-L.}\ \bibnamefont {Barab{\'a}si}},\
  }\href@noop {} {\bibfield  {journal} {\bibinfo  {journal} {Science}\ }\textbf
  {\bibinfo {volume} {342}},\ \bibinfo {pages} {127} (\bibinfo {year}
  {2013})}\BibitemShut {NoStop}%
\bibitem [{\citenamefont {Redner}(2005)}]{redner2005citation}%
  \BibitemOpen
  \bibfield  {author} {\bibinfo {author} {\bibfnamefont {S.}~\bibnamefont
  {Redner}},\ }\href@noop {} {\bibfield  {journal} {\bibinfo  {journal} {arXiv
  preprint physics/0506056}\ } (\bibinfo {year} {2005})}\BibitemShut {NoStop}%
\bibitem [{\citenamefont {Redner}(1998)}]{redner1998popular}%
  \BibitemOpen
  \bibfield  {author} {\bibinfo {author} {\bibfnamefont {S.}~\bibnamefont
  {Redner}},\ }\href@noop {} {\bibfield  {journal} {\bibinfo  {journal} {The
  European Physical Journal B-Condensed Matter and Complex Systems}\ }\textbf
  {\bibinfo {volume} {4}},\ \bibinfo {pages} {131} (\bibinfo {year}
  {1998})}\BibitemShut {NoStop}%
\bibitem [{\citenamefont {Uzzi}\ \emph {et~al.}(2013)\citenamefont {Uzzi},
  \citenamefont {Mukherjee}, \citenamefont {Stringer},\ and\ \citenamefont
  {Jones}}]{uzzi2013atypical}%
  \BibitemOpen
  \bibfield  {author} {\bibinfo {author} {\bibfnamefont {B.}~\bibnamefont
  {Uzzi}}, \bibinfo {author} {\bibfnamefont {S.}~\bibnamefont {Mukherjee}},
  \bibinfo {author} {\bibfnamefont {M.}~\bibnamefont {Stringer}}, \ and\
  \bibinfo {author} {\bibfnamefont {B.}~\bibnamefont {Jones}},\ }\href@noop {}
  {\bibfield  {journal} {\bibinfo  {journal} {Science}\ }\textbf {\bibinfo
  {volume} {342}},\ \bibinfo {pages} {468} (\bibinfo {year}
  {2013})}\BibitemShut {NoStop}%
\bibitem [{\citenamefont {Yin}\ and\ \citenamefont {Wang}(2017)}]{yin2017time}%
  \BibitemOpen
  \bibfield  {author} {\bibinfo {author} {\bibfnamefont {Y.}~\bibnamefont
  {Yin}}\ and\ \bibinfo {author} {\bibfnamefont {D.}~\bibnamefont {Wang}},\
  }\href@noop {} {\bibfield  {journal} {\bibinfo  {journal} {Journal of
  Informetrics}\ }\textbf {\bibinfo {volume} {11}},\ \bibinfo {pages} {608}
  (\bibinfo {year} {2017})}\BibitemShut {NoStop}%
\bibitem [{\citenamefont {Golosovsky}(2017)}]{golosovsky2017power}%
  \BibitemOpen
  \bibfield  {author} {\bibinfo {author} {\bibfnamefont {M.}~\bibnamefont
  {Golosovsky}},\ }\href@noop {} {\bibfield  {journal} {\bibinfo  {journal}
  {Physical Review E}\ }\textbf {\bibinfo {volume} {96}},\ \bibinfo {pages}
  {032306} (\bibinfo {year} {2017})}\BibitemShut {NoStop}%
\bibitem [{\citenamefont {Stringer}, \citenamefont {Sales-Pardo},\ and\
  \citenamefont {Amaral}(2010)}]{stringer2010statistical}%
  \BibitemOpen
  \bibfield  {author} {\bibinfo {author} {\bibfnamefont {M.~J.}\ \bibnamefont
  {Stringer}}, \bibinfo {author} {\bibfnamefont {M.}~\bibnamefont
  {Sales-Pardo}}, \ and\ \bibinfo {author} {\bibfnamefont {L.~A.~N.}\
  \bibnamefont {Amaral}},\ }\href@noop {} {\bibfield  {journal} {\bibinfo
  {journal} {Journal of the American Society for Information Science and
  Technology}\ }\textbf {\bibinfo {volume} {61}},\ \bibinfo {pages} {1377}
  (\bibinfo {year} {2010})}\BibitemShut {NoStop}%
\bibitem [{\citenamefont {Barab{\'a}si}, \citenamefont {Song},\ and\
  \citenamefont {Wang}(2012)}]{barabasi2012handful}%
  \BibitemOpen
  \bibfield  {author} {\bibinfo {author} {\bibfnamefont {A.-L.}\ \bibnamefont
  {Barab{\'a}si}}, \bibinfo {author} {\bibfnamefont {C.}~\bibnamefont {Song}},
  \ and\ \bibinfo {author} {\bibfnamefont {D.}~\bibnamefont {Wang}},\
  }\href@noop {} {\bibfield  {journal} {\bibinfo  {journal} {Nature}\ }\textbf
  {\bibinfo {volume} {491}},\ \bibinfo {pages} {40} (\bibinfo {year}
  {2012})}\BibitemShut {NoStop}%
\bibitem [{ISI(2012)}]{ISI}%
  \BibitemOpen
  \href@noop {} {\enquote {\bibinfo {title} {{ISI Web of Knowledge}},}\
  }\bibinfo {howpublished} {\url{http://scientific.thomson.com/isi/}} (\bibinfo
  {year} {2012}),\ \bibinfo {note} {[accessed January 2012]}\BibitemShut
  {NoStop}%
\bibitem [{\citenamefont {Bouchaud}(1992)}]{bouchaud1992weak}%
  \BibitemOpen
  \bibfield  {author} {\bibinfo {author} {\bibfnamefont {J.-P.}\ \bibnamefont
  {Bouchaud}},\ }\href@noop {} {\bibfield  {journal} {\bibinfo  {journal}
  {Journal de Physique I}\ }\textbf {\bibinfo {volume} {2}},\ \bibinfo {pages}
  {1705} (\bibinfo {year} {1992})}\BibitemShut {NoStop}%
\bibitem [{\citenamefont {Bel}\ and\ \citenamefont
  {Barkai}(2005)}]{bel2005weak}%
  \BibitemOpen
  \bibfield  {author} {\bibinfo {author} {\bibfnamefont {G.}~\bibnamefont
  {Bel}}\ and\ \bibinfo {author} {\bibfnamefont {E.}~\bibnamefont {Barkai}},\
  }\href@noop {} {\bibfield  {journal} {\bibinfo  {journal} {Physical review
  letters}\ }\textbf {\bibinfo {volume} {94}},\ \bibinfo {pages} {240602}
  (\bibinfo {year} {2005})}\BibitemShut {NoStop}%
\bibitem [{\citenamefont {Burov}\ \emph {et~al.}(2011)\citenamefont {Burov},
  \citenamefont {Jeon}, \citenamefont {Metzler},\ and\ \citenamefont
  {Barkai}}]{burov2011single}%
  \BibitemOpen
  \bibfield  {author} {\bibinfo {author} {\bibfnamefont {S.}~\bibnamefont
  {Burov}}, \bibinfo {author} {\bibfnamefont {J.-H.}\ \bibnamefont {Jeon}},
  \bibinfo {author} {\bibfnamefont {R.}~\bibnamefont {Metzler}}, \ and\
  \bibinfo {author} {\bibfnamefont {E.}~\bibnamefont {Barkai}},\ }\href@noop {}
  {\bibfield  {journal} {\bibinfo  {journal} {Physical Chemistry Chemical
  Physics}\ }\textbf {\bibinfo {volume} {13}},\ \bibinfo {pages} {1800}
  (\bibinfo {year} {2011})}\BibitemShut {NoStop}%
\bibitem [{\citenamefont {Cherstvy}, \citenamefont {Chechkin},\ and\
  \citenamefont {Metzler}(2013)}]{cherstvy2013anomalous}%
  \BibitemOpen
  \bibfield  {author} {\bibinfo {author} {\bibfnamefont {A.~G.}\ \bibnamefont
  {Cherstvy}}, \bibinfo {author} {\bibfnamefont {A.~V.}\ \bibnamefont
  {Chechkin}}, \ and\ \bibinfo {author} {\bibfnamefont {R.}~\bibnamefont
  {Metzler}},\ }\href@noop {} {\bibfield  {journal} {\bibinfo  {journal} {New
  Journal of Physics}\ }\textbf {\bibinfo {volume} {15}},\ \bibinfo {pages}
  {083039} (\bibinfo {year} {2013})}\BibitemShut {NoStop}%
\bibitem [{\citenamefont {Thiel}\ and\ \citenamefont
  {Sokolov}(2014)}]{thiel2014weak}%
  \BibitemOpen
  \bibfield  {author} {\bibinfo {author} {\bibfnamefont {F.}~\bibnamefont
  {Thiel}}\ and\ \bibinfo {author} {\bibfnamefont {I.~M.}\ \bibnamefont
  {Sokolov}},\ }\href@noop {} {\bibfield  {journal} {\bibinfo  {journal}
  {Physical Review E}\ }\textbf {\bibinfo {volume} {89}},\ \bibinfo {pages}
  {012136} (\bibinfo {year} {2014})}\BibitemShut {NoStop}%
\bibitem [{\citenamefont {Metzler}(2015)}]{metzler2015weak}%
  \BibitemOpen
  \bibfield  {author} {\bibinfo {author} {\bibfnamefont {R.}~\bibnamefont
  {Metzler}},\ }in\ \href@noop {} {\emph {\bibinfo {booktitle} {International
  Journal of Modern Physics: Conference Series}}},\ Vol.~\bibinfo {volume}
  {36}\ (\bibinfo {organization} {World Scientific},\ \bibinfo {year} {2015})\
  p.\ \bibinfo {pages} {1560007}\BibitemShut {NoStop}%
\bibitem [{\citenamefont {Manzo}\ \emph {et~al.}(2015)\citenamefont {Manzo},
  \citenamefont {Torreno-Pina}, \citenamefont {Massignan}, \citenamefont
  {Lapeyre~Jr}, \citenamefont {Lewenstein},\ and\ \citenamefont
  {Parajo}}]{manzo2015weak}%
  \BibitemOpen
  \bibfield  {author} {\bibinfo {author} {\bibfnamefont {C.}~\bibnamefont
  {Manzo}}, \bibinfo {author} {\bibfnamefont {J.~A.}\ \bibnamefont
  {Torreno-Pina}}, \bibinfo {author} {\bibfnamefont {P.}~\bibnamefont
  {Massignan}}, \bibinfo {author} {\bibfnamefont {G.~J.}\ \bibnamefont
  {Lapeyre~Jr}}, \bibinfo {author} {\bibfnamefont {M.}~\bibnamefont
  {Lewenstein}}, \ and\ \bibinfo {author} {\bibfnamefont {M.~F.~G.}\
  \bibnamefont {Parajo}},\ }\href@noop {} {\bibfield  {journal} {\bibinfo
  {journal} {Physical Review X}\ }\textbf {\bibinfo {volume} {5}},\ \bibinfo
  {pages} {011021} (\bibinfo {year} {2015})}\BibitemShut {NoStop}%
\bibitem [{\citenamefont {Mandelbrot}\ and\ \citenamefont
  {Wallis}(1968)}]{mandelbrot1968noah}%
  \BibitemOpen
  \bibfield  {author} {\bibinfo {author} {\bibfnamefont {B.~B.}\ \bibnamefont
  {Mandelbrot}}\ and\ \bibinfo {author} {\bibfnamefont {J.~R.}\ \bibnamefont
  {Wallis}},\ }\href@noop {} {\bibfield  {journal} {\bibinfo  {journal} {Water
  resources research}\ }\textbf {\bibinfo {volume} {4}},\ \bibinfo {pages}
  {909} (\bibinfo {year} {1968})}\BibitemShut {NoStop}%
\bibitem [{\citenamefont {Chen}\ \emph {et~al.}(2017)\citenamefont {Chen},
  \citenamefont {Bassler}, \citenamefont {McCauley},\ and\ \citenamefont
  {Gunaratne}}]{chen2017anomalous}%
  \BibitemOpen
  \bibfield  {author} {\bibinfo {author} {\bibfnamefont {L.}~\bibnamefont
  {Chen}}, \bibinfo {author} {\bibfnamefont {K.~E.}\ \bibnamefont {Bassler}},
  \bibinfo {author} {\bibfnamefont {J.~L.}\ \bibnamefont {McCauley}}, \ and\
  \bibinfo {author} {\bibfnamefont {G.~H.}\ \bibnamefont {Gunaratne}},\
  }\href@noop {} {\bibfield  {journal} {\bibinfo  {journal} {Physical Review
  E}\ }\textbf {\bibinfo {volume} {95}},\ \bibinfo {pages} {042141} (\bibinfo
  {year} {2017})}\BibitemShut {NoStop}%
\bibitem [{\citenamefont {Meyer}\ \emph {et~al.}(2018)\citenamefont {Meyer},
  \citenamefont {Adlakha}, \citenamefont {Kantz},\ and\ \citenamefont
  {Bassler}}]{meyer2018anomalous}%
  \BibitemOpen
  \bibfield  {author} {\bibinfo {author} {\bibfnamefont {P.~G.}\ \bibnamefont
  {Meyer}}, \bibinfo {author} {\bibfnamefont {V.}~\bibnamefont {Adlakha}},
  \bibinfo {author} {\bibfnamefont {H.}~\bibnamefont {Kantz}}, \ and\ \bibinfo
  {author} {\bibfnamefont {K.~E.}\ \bibnamefont {Bassler}},\ }\href@noop {}
  {\bibfield  {journal} {\bibinfo  {journal} {New Journal of Physics}\ }\textbf
  {\bibinfo {volume} {20}},\ \bibinfo {pages} {113033} (\bibinfo {year}
  {2018})}\BibitemShut {NoStop}%
\bibitem [{\citenamefont {Aghion}\ \emph {et~al.}(2020)\citenamefont {Aghion},
  \citenamefont {Meyer}, \citenamefont {Adlakha}, \citenamefont {Kantz},\ and\
  \citenamefont {Bassler}}]{aghion2020moses}%
  \BibitemOpen
  \bibfield  {author} {\bibinfo {author} {\bibfnamefont {E.}~\bibnamefont
  {Aghion}}, \bibinfo {author} {\bibfnamefont {P.~G.}\ \bibnamefont {Meyer}},
  \bibinfo {author} {\bibfnamefont {V.}~\bibnamefont {Adlakha}}, \bibinfo
  {author} {\bibfnamefont {H.}~\bibnamefont {Kantz}}, \ and\ \bibinfo {author}
  {\bibfnamefont {K.}~\bibnamefont {Bassler}},\ }\href@noop {} {\bibfield
  {journal} {\bibinfo  {journal} {New Journal of Physics}\ } (\bibinfo {year}
  {2020})}\BibitemShut {NoStop}%
\bibitem [{\citenamefont {Clauset}, \citenamefont {Shalizi},\ and\
  \citenamefont {Newman}(2009)}]{clauset2009power}%
  \BibitemOpen
  \bibfield  {author} {\bibinfo {author} {\bibfnamefont {A.}~\bibnamefont
  {Clauset}}, \bibinfo {author} {\bibfnamefont {C.~R.}\ \bibnamefont
  {Shalizi}}, \ and\ \bibinfo {author} {\bibfnamefont {M.~E.}\ \bibnamefont
  {Newman}},\ }\href@noop {} {\bibfield  {journal} {\bibinfo  {journal} {SIAM
  review}\ }\textbf {\bibinfo {volume} {51}},\ \bibinfo {pages} {661} (\bibinfo
  {year} {2009})}\BibitemShut {NoStop}%
\bibitem [{\citenamefont {Brzezinski}(2015)}]{brzezinski2015power}%
  \BibitemOpen
  \bibfield  {author} {\bibinfo {author} {\bibfnamefont {M.}~\bibnamefont
  {Brzezinski}},\ }\href@noop {} {\bibfield  {journal} {\bibinfo  {journal}
  {Scientometrics}\ }\textbf {\bibinfo {volume} {103}},\ \bibinfo {pages} {213}
  (\bibinfo {year} {2015})}\BibitemShut {NoStop}%
\bibitem [{\citenamefont {Eom}\ and\ \citenamefont
  {Fortunato}(2011)}]{eom2011characterizing}%
  \BibitemOpen
  \bibfield  {author} {\bibinfo {author} {\bibfnamefont {Y.-H.}\ \bibnamefont
  {Eom}}\ and\ \bibinfo {author} {\bibfnamefont {S.}~\bibnamefont
  {Fortunato}},\ }\href@noop {} {\bibfield  {journal} {\bibinfo  {journal}
  {PloS one}\ }\textbf {\bibinfo {volume} {6}},\ \bibinfo {pages} {e24926}
  (\bibinfo {year} {2011})}\BibitemShut {NoStop}%
\bibitem [{\citenamefont {Evans}, \citenamefont {Hopkins},\ and\ \citenamefont
  {Kaube}(2012)}]{evans2012universality}%
  \BibitemOpen
  \bibfield  {author} {\bibinfo {author} {\bibfnamefont {T.}~\bibnamefont
  {Evans}}, \bibinfo {author} {\bibfnamefont {N.}~\bibnamefont {Hopkins}}, \
  and\ \bibinfo {author} {\bibfnamefont {B.}~\bibnamefont {Kaube}},\
  }\href@noop {} {\bibfield  {journal} {\bibinfo  {journal} {Scientometrics}\
  }\textbf {\bibinfo {volume} {93}},\ \bibinfo {pages} {473} (\bibinfo {year}
  {2012})}\BibitemShut {NoStop}%
\bibitem [{\citenamefont {Bianconi}\ and\ \citenamefont
  {Barab{\'a}si}(2001)}]{bianconi2001bose}%
  \BibitemOpen
  \bibfield  {author} {\bibinfo {author} {\bibfnamefont {G.}~\bibnamefont
  {Bianconi}}\ and\ \bibinfo {author} {\bibfnamefont {A.-L.}\ \bibnamefont
  {Barab{\'a}si}},\ }\href@noop {} {\bibfield  {journal} {\bibinfo  {journal}
  {Physical review letters}\ }\textbf {\bibinfo {volume} {86}},\ \bibinfo
  {pages} {5632} (\bibinfo {year} {2001})}\BibitemShut {NoStop}%
\bibitem [{\citenamefont {Albert}\ and\ \citenamefont
  {Barab{\'a}si}(2002)}]{albert2002statistical}%
  \BibitemOpen
  \bibfield  {author} {\bibinfo {author} {\bibfnamefont {R.}~\bibnamefont
  {Albert}}\ and\ \bibinfo {author} {\bibfnamefont {A.-L.}\ \bibnamefont
  {Barab{\'a}si}},\ }\href@noop {} {\bibfield  {journal} {\bibinfo  {journal}
  {Reviews of modern physics}\ }\textbf {\bibinfo {volume} {74}},\ \bibinfo
  {pages} {47} (\bibinfo {year} {2002})}\BibitemShut {NoStop}%
\bibitem [{\citenamefont {Krapivsky}\ and\ \citenamefont
  {Redner}(2001)}]{krapivsky2001organization}%
  \BibitemOpen
  \bibfield  {author} {\bibinfo {author} {\bibfnamefont {P.~L.}\ \bibnamefont
  {Krapivsky}}\ and\ \bibinfo {author} {\bibfnamefont {S.}~\bibnamefont
  {Redner}},\ }\href@noop {} {\bibfield  {journal} {\bibinfo  {journal}
  {Physical Review E}\ }\textbf {\bibinfo {volume} {63}},\ \bibinfo {pages}
  {066123} (\bibinfo {year} {2001})}\BibitemShut {NoStop}%
\bibitem [{\citenamefont {Klafter}\ and\ \citenamefont
  {Sokolov}(2011)}]{klafter2011first}%
  \BibitemOpen
  \bibfield  {author} {\bibinfo {author} {\bibfnamefont {J.}~\bibnamefont
  {Klafter}}\ and\ \bibinfo {author} {\bibfnamefont {I.~M.}\ \bibnamefont
  {Sokolov}},\ }\href@noop {} {\emph {\bibinfo {title} {First steps in random
  walks: from tools to applications}}}\ (\bibinfo  {publisher} {Oxford
  University Press},\ \bibinfo {year} {2011})\BibitemShut {NoStop}%
\bibitem [{\citenamefont {Lim}\ and\ \citenamefont
  {Muniandy}(2002)}]{lim2002self}%
  \BibitemOpen
  \bibfield  {author} {\bibinfo {author} {\bibfnamefont {S.}~\bibnamefont
  {Lim}}\ and\ \bibinfo {author} {\bibfnamefont {S.}~\bibnamefont {Muniandy}},\
  }\href@noop {} {\bibfield  {journal} {\bibinfo  {journal} {Physical Review
  E}\ }\textbf {\bibinfo {volume} {66}},\ \bibinfo {pages} {021114} (\bibinfo
  {year} {2002})}\BibitemShut {NoStop}%
\bibitem [{\citenamefont {Bormashenko}(2019)}]{bormashenko2019moses}%
  \BibitemOpen
  \bibfield  {author} {\bibinfo {author} {\bibfnamefont {E.}~\bibnamefont
  {Bormashenko}},\ }\href@noop {} {\bibfield  {journal} {\bibinfo  {journal}
  {Advances in colloid and interface science}\ }\textbf {\bibinfo {volume}
  {269}},\ \bibinfo {pages} {1} (\bibinfo {year} {2019})}\BibitemShut {NoStop}%
\bibitem [{\citenamefont {Mandelbrot}(2002)}]{mandelbrot2002gaussian}%
  \BibitemOpen
  \bibfield  {author} {\bibinfo {author} {\bibfnamefont {B.}~\bibnamefont
  {Mandelbrot}},\ }\href@noop {} {\emph {\bibinfo {title} {Gaussian
  self-affinity and fractals: globality, the earth, 1/f noise, and R/S}}},\
  Vol.~\bibinfo {volume} {8}\ (\bibinfo  {publisher} {Springer Science \&
  Business Media},\ \bibinfo {year} {2002})\BibitemShut {NoStop}%
\end{thebibliography}%

\end{document}